  \documentclass[preprint2]{aastex7}

\def\dsm{$\mathrm{M}_\odot$}

\def\teff{$T_{\mathrm{eff}}$}
\def\alph{$\alpha_{\rm{MLT}}$}
\def\ali{$A$(Li)}

\def\tbc{$T_{\mathrm{BCZ}}$}

\begin{document}

\title{Using Lithium and Beryllium to Study Structure and Evolution of Rotating Stars:
Spite Plateau of Halo Stars}

\author[orcid=0000-0002-3956-8061, gname=Wuming, sname=Yang]{Wuming Yang}
\affiliation{School of Physics and Astronomy, Beijing Normal University, Beijing 100875, China}
\email[show]{yangwuming@bnu.edu.cn}

\author[gname=Shuya,sname=Dou]{Shuya Dou}
\affiliation{School of Physics and Astronomy, Beijing Normal University, Beijing 100875, China}
\email{202521101099@mail.bnu.edu.cn}

\author[gname=Xiangcun,sname=Meng]{Xiangcun Meng}
\affiliation{International Centre of Supernovae, Yunnan Key Laboratory of Supernova Research,
Yunnan Observatories, Chinese Academy of Sciences, Kunming 650216, China}
\email{xiangcunmeng@ynao.ac.cn}

\author[gname=Yaqian, sname=Wu]{Yaqian Wu}
\affiliation{Key Laboratory of Optical Astronomy, National Astronomical Observatories,
Chinese Academy of Sciences, A20 Datun Road, Chaoyang District, Beijing, 100101, China}
\email{wuyaqian@nao.cas.cn}

\author[gname=Shaolan, sname=Bi]{Shaolan Bi}
\affiliation{School of Physics and Astronomy, Beijing Normal University, Beijing 100875, China}
\email{bisl@bnu.edu.cn}
\begin{abstract}
The observed lithum (Li) abundance of Galactic halo stars mainly fall within the range of 2.0--2.4 dex.
This nearly constant value, known as the Spite plateau, is approximately a factor of three lower than
the value predicted from cosmic microwave background measurements and standard Big Bang Nucleosynthesis
(BBN) calculations. This discrepancy---referred to as the cosmological Li problem---is considered a potential
indication of new physics or astrophysical processes. We employed models incorporating gravitational
settling, diffusion, rotation, and magnetic fields to explain the Spite plateau. The rotating models
predict that Li abundances in stars with ages of roughly 8--13 Gyr and effective temperatures
between 6400 and 5900 K generally fall within 2.0--2.4 dex, forming a well-defined
Li plateau, followed by a sharp decline in Li abundance down to about 5200 K. The Li plateau results from
the combined effects of variations in convection zone depth, gravitational settling, diffusion, rotation,
and magnetic fields. For red giant branch stars with $T_{\mathrm{eff}} \lesssim$ 5200 K, the rotating models
predict another Li plateau with an abundance of about 1.0 dex. These results are in good agreement with observations.
Moreover, the initial Li abundance of 2.72 dex adopted in the models matches the BBN prediction, implying
that the Li problem arises from stellar Li depletion. Furthermore, the rotating models also reproduce
the Li and Be distributions of the sample that exhibit the Spite plateau meltdown and Be deviation.
\end{abstract}

\section{Introduction}
Lithium-7 ($^{7}$Li) and beryllium-9 ($^{9}$Be) in stars are easily destroyed by energetic
protons at temperatures near 2.5 $\times$ $10^{6}$ K and 3.5 $\times$ $10^{6}$ K, respectively, making them fragile
elements. The Sun's initial Li and Be abundances are \ali{} = 3.3 dex \citep{lodd21} and $A(\mathrm{Be})$ = 1.44
dex \citep{aspl21}, respectively, where $A(\mathrm{x})\equiv\log(N(\mathrm{x})/N(\mathrm{H}))$ + 12. In contrast,
the present-day Li and Be abundances in the solar photosphere---and thus in the convection zone (CZ)---are 0.96
$\pm$ 0.05 dex \citep{wang21} and 1.32 $\pm$ 0.05 dex \citep{koro22}, respectively. Solar models predict that the
temperature at the base of the CZ (BCZ) is about 2.2 $\times$ $10^{6}$ K \citep{yang25}. Therefore, the depletion
of $^{7}$Li and $^{9}$Be in the Sun is closely linked to the internal stellar structure and the physical processes
occurring within and beneath the CZ. Both Li and Be serve as excellent astrophysical tracers of transport and mixing
processes in these regions. To explain both the seismically inferred rotation profile and the observed solar Li and
Be abundances, transport and mixing processes driven by hydrodynamic and magnetic instabilities are required
\citep{egge22, yang25}.

Lithium-7 is one of the four primordial isotopes ($^{2}$H, $^{3}$He, $^{4}$He, and $^{7}$Li) formed during
the first $\sim$15 minutes of the Big Bang, when the universe was dense and hot enough for nuclear reactions
to occur. Its primordial abundance, predicted by the standard Big Bang Nucleosynthesis (BBN), mainly depends
on the baryon-to-photon ratio $\eta$, which is proportional to $\Omega_{b}h^{2}$, where $\Omega_{b}$ is the baryonic
matter density parameter in cosmology and $h$ is the Hubble parameter \citep{coc12, coc17}. Based on Wilkinson
Microwave Anisotropy Probe (WMAP; \citealt{benn03}) and Planck \citep{plan14} data, BBN predicts
$N(\mathrm{^{7}Li})/N(\mathrm{H})$ = (5.24$^{+0.71}_{-0.67}$) $\times\ 10^{-10}$, corresponding to $A(\mathrm{Li})$
= 2.72 $\pm$ 0.06 \citep{cybu08, coc14, sing24}.

In classical stellar models that neglect gravitational settling, rotational mixing, and magnetic fields, the surface
Li abundance of a star with a shallow CZ remains nearly constant. The lifetime of a star with $M \lesssim 0.8$
\dsm{} is expected to exceed the age of the universe (13.8 Gyr; \citealt{plan16}). Therefore, the primordial Li
abundance could be preserved at the surface of the oldest, first-generation stars. For near-main-sequence (MS) stars
with 5700 K $<$ \teff{} $<$ 6500 K and $-2.5 \lesssim$ [Fe/H] $\lesssim -1$ in the Galactic halo, the observed Li
abundance lies mainly in the range of 2.0--2.4 dex \citep{spit82, charb05, sbor10}. This nearly constant value,
known as the \textit{Spite Plateau}, is about a factor of three lower than the value predicted from cosmic microwave
background (CMB) and BBN determinations. Moreover, no significant dispersion has been observed along the plateau
\citep{charb05, aspl06}. This discrepancy is referred to as \textit{the cosmological lithium problem}.

Similar results have been reported in metal-poor globular clusters, including M92 \citep{boes98}, NGC 6397 \citep{korn06,
korn07, lind09}, M5 \citep{boes23}, M13, and M71 \citep{boes24}, as well as in extragalactic systems \citep{mola20, matt21}.
In NGC 6397, \citet{korn06, korn07} reported a mean Li abundance of 2.24 $\pm$ 0.05 for five turn-off (TO) stars and
2.36 $\pm$ 0.05 for two subgiant branch (SGB) stars. These findings were later confirmed by \citet{lind09},
who likewise found that TO stars are more Li-poor than subgiants that have not yet undergone dredge-up. Moreover,
no significant Li dispersion was detected along the plateau of NGC 6397. However, a significant anti-correlation
between Li and Na abundances was observed in the SGB stars of this cluster \citep{lind09}.

The presence of dispersion in \ali{} remains a subject of debate. A dependence of \ali{} on both \teff{} and
metallicity was reported by \citet{deli93}, \citet{norr94}, \citet{thor94}, \citet{gao20}, and \citet{roma21},
but was challenged by \citet{mola95} and \citet{spit96}. In addition, \citet{boni97} found a slight trend with \teff{}
but no correlation with [Fe/H], whereas \citet{ryan99} and \citet{aspl06} reported a small positive slope of \ali{}
with [Fe/H]. Further research is required to explain these controversial results.

Moreover, \citet{sbor10} found evidence for a meltdown of the Spite plateau in stars with [Fe/H] $\lesssim -3$.
In contrast, \citet{smil09, smil21} reported deviations in Be abundances for stars with [Fe/H] around $-3$.
The origin of these phenomena remains unknown.

Primordial nucleosynthesis represents one of the most fundamental processes in cosmic history, for which all
relevant physics is known a priori. Consequently, the discrepancy between the Li abundances observed in metal-poor
halo stars and those predicted by CMB + BBN is regarded as a potential indication of new physics or astrophysical
processes \citep{iocc09, field11, coc17, deal21}. Considerable research has been devoted to resolving the Li problem,
and nuclear-physics solutions have largely been ruled out \citep{hamm13, coc17}. Instead, the discrepancy may point
toward physics beyond the Standard Model of particle physics and/or standard cosmology \citep{field11, clar20, deal21,
groh22}. \citet{agua19} argued that the primordial Li abundance may be lower than generally accepted. Thus,
the possibility of beyond–Standard Model physics remains open.

Furthermore, numerous studies attribute the origin of the Li problem to stellar Li depletion processes,
incorporating the effects of gravitational settling \citep{mich84, rich05}, rotational mixing \citep{deal21, bori24},
internal gravity waves \citep{talo04}, mass loss \citep{vick13}, convective overshooting, and residual mass
accretion \citep{fu15}.

\citet{vick13} concluded that reproducing plateau-like Li abundances in Population II
stars requires a high mass-loss rate, although such a rate appears implausible when compared with the Sun.
\citet{deal21} demonstrated that the combined effects of atomic diffusion, rotation, and penetrative convection
can account for the Li problem. However, \citet{bori24} showed that rotation-induced mixing alone is insufficient
to reproduce the observed Li distribution, and that additional parametric turbulence is necessary. Moreover,
\citet{nguy25a} showed that the Spite plateau can be explained by the convective overshoot depending on stellar mass
and evolutionary stage. Thus, the underlying nature of the Li problem remains unresolved and continues to be
the subject of considerable debate.

In this work, we investigate whether the cosmological lithium problem can be explained by the same mechanisms
responsible for solar lithium depletion---namely, gravitational settling, diffusion, rotation, and magnetic fields.
The organization of this paper is as follows. Section \ref{sec2} describes the input physics,
Section \ref{sec3} presents the calculation results, and Section \ref{sec4} provides the
discussion and conclusions.

\section{Evolution Code and Input Physics} \label{sec2}

We employ the Yale Rotating Stellar Evolution Code (YREC, \citealt{enda76, enda78, pins89, yang07, yang16, yang25}) to compute
stellar evolutionary models. The OPAL equation-of-state (EOS2005) tables \citep{roge02} and OPAL opacity tables \citep{igle96}
are adopted, supplemented by the low-temperature opacity tables of \citet{ferg05}. These opacity tables are reconstructed
according to the mixtures of \citet{magg22}.

In regions with $2\times10^{6}$ K $\lesssim T \lesssim 5\times10^{6}$ K, the Rosseland mean opacity is increased linearly by
no more than $2.5\%$ centered at $T=3\times10^{6}$ K \citep{yang22, yang24}. Gravitational settling and diffusion of both helium
and heavy elements are included using the diffusion coefficients of \citet{thou94}, while the effects of radiative levitation
on chemical transport \citep{turc98} are not incorporated. The impact of radiative acceleration may be mitigated
by the effects of rotation and magnetic ﬁelds.

Since the concentrations of Li, Be, and boron (B) are too low, they are generally not included in the total metal
abundance $Z$ and are not considered in the diffusion and settling of heavy elements. In this work, however, we
incorporated the diffusion and settling of Li, Be, and B into all models.

For the atmosphere, the \citet{kris66} $T-\tau$ relation is adopted. The boundary of the CZ is determined by the
Schwarzschild criterion, and energy transfer by convection is treated according to the standard mixing-length
theory (MLT) \citep{bohm58}. The depth of the overshoot region, if present, is given by $\delta_{\mathrm{ov}}H_{p}$,
where $\delta_{\mathrm{ov}}$ is a free parameter, and $H_{p}$ is the local pressure scale height. The overshoot region
is assumed to be adiabatically stratiﬁed and fully mixed.

Nuclear reaction rates are calculated using the subroutines of \citet{bahc92}, updated by \citet{bahc95, bahc01}
and \citet{yang24}. The burning rates of Li, Be, and B are computed as a function of temperature for $T \geq10^{6}$ K,
under the assumption that these elements are completely destroyed at $T>10^{7}$ K.

The impacts of disc-locking during the early pre-MS phase and magnetic braking are incorporated \citep{yang25}.
During the disc-locking phase, we simply assume that the angular velocity of the stellar surface CZ remains constant,
with a disk-locking timescale of 5 Myr. Angular momentum loss from the CZ due to magnetic braking is calculated using
Kawaler’s relation \citep{kawa88, chab95}.

The internal angular-momentum transport and chemical composition mixing in radiative zones are treated as
diffusion processes \citep{enda78, yang06}
    \begin{equation}
       \frac{\partial \Omega}{\partial t}=
       \frac{1}{\rho r^{4}}\frac{\partial}{\partial r}[(\rho r^{4}(f_{\Omega}D_{r}+f_{m}D_{m})
       \frac{\partial \Omega}{\partial r})],
      \label{diffu1}
    \end{equation}
for angular momentum transport and
    \begin{equation}
    \begin{array}{lll}
        \frac{\partial X_{i}}{\partial t}&=&\frac{1}{\rho r^{2}}
       \frac{\partial}{\partial r}[\rho r^{2}(f_{c}f_{\Omega}D_{r}+f_{cm}f_{m}D_{m})\frac{\partial X_{i}}
        {\partial r}]\\
        & &+(\frac{\partial X_{i}}{\partial t})_{\rm nuc}-\frac{1}
       {\rho r^{2}}\frac{\partial}{\partial r}(\rho r^{2}X_{i}V_{i}),
    \end{array}
      \label{diffu2}
    \end{equation}
for the change in the mass fraction $X_{i}$ of chemical species $i$, where $\rho$ is the density, $V_{i}$
is the velocity of microscopic diffusion and settling given by \citet{thou94}, and $D_{r}$ is the diffusion
coefficient associated with rotational instabilities. These instabilities include the dynamical instabilities
described in \citet{enda78} and \citet{pins89}, as well as the secular shear instability \citep{zahn93}:
\begin{equation}
 D = \frac{2c}{27G}|\frac{d\ln T}{dr}-\frac{2}{3}
       \frac{d\ln\rho}{dr}|^{-1}\frac{r^{4}}{\kappa \rho M(r)}
       (\frac{d\Omega}{dr})^{2},
\end{equation}
where $\kappa$ is the Rosseland mean opacity, $c$ is the speed of light, and $G$ is gravitational constant.
The rotational mixing and diffusion are treated separately in YREC.

The diffusion coefficient induced by magnetic fields is defined as \citep{yang06}
\begin{equation}
 D_{m}=r^{2}\Omega\frac{B^{2}_{r}}{B^{2}},
\end{equation}
where the magnetic ﬁeld compositions are calculated using Equations (22) and (23) of \citet{spru02}.
The parameters $f_{\Omega}$ and $f_{m}$ are introduced to account for uncertainties in the diffusion equations,
while $f_{c}$ and $f_{cm}$ reflect the fact that the instabilities and magnetic fields mix chemical material
less efficiently than they transport angular momentum \citep{pins89, yang16}. The larger the values
of $f_{c}$ or $f_{cm}$, the higher the efficiency of the mixing \citep{yang25}. The default parameter values
are $f_{\Omega}=1$, $f_{c} = 0.03$, $f_{m}=0.0001$, and $f_{cm}=0.0002$. These parameters are calibrated to
reproduce the seismically inferred solar rotation profile and surface helium abundance, as well as the observed
solar Li and Be abundances \citep{yang25}. Below the BCZ of rotating models (RMs), a tachocline with a width
of $0.05 R$ (where $R$ is the stellar radius; \citealt{charb99}) is assumed, following \citet{yang25}.

The initial rotation rate, $\Omega_{i}$, of models is treated as a free parameter. The stars observed in M13
and M71 by \citet{boes24} rotate so slowly that $v\sin i$ cannot be measured from their spectra. The solar model
of \citet{yang25}, which reproduces helioseismic results, neutrino fluxes, and the observed Li and Be abundances,
is also a slow rotator, with a zero-age main sequence (ZAMS) rotation rate of about 3.6 $\Omega_{\odot}$,
corresponding to $\Omega_{i} \simeq 2\times10^{-6}$ rad s$^{-1}$ at the initial time. In this work, we adopt this
value as one of initial rotation rates of RMs.

The initial hydrogen abundance $X_{0}$, metal abundance $Z_{0}$, and mixing-length parameter \alph{} are
also treated as free parameters in YREC. The value of \alph{} is calibrated to the Sun \citep{yang25}
and assumed to remain constant. In our calculations, the initial hydrogen abundance is given by
\begin{equation}
 X_{0}=0.762-3Z_{0},
 \label{xi}
\end{equation}
which corresponds to an initial helium abundance of
\begin{equation}
 Y_{0}=0.238+2Z_{0},
 \label{yi}
\end{equation}
where the primordial helium abundance of 0.238 is inferred from observations of metal-poor H II regions
\citep{fiel98, luri03}. This value, however, is lower than the 0.24705 $\pm$ 0.00019 predicted by
\citet{pitr18}. The helium-to-metal enrichment ratio, $\Delta Y/\Delta Z$ = 2, is determined from
the primordial helium abundance and the solar values of $Y_{0}$ and $Z_{0}$ \citep{yang25}. Thus,
equations (\ref{xi}) and (\ref{yi}) are constrained by both observations of metal-poor H II regions
and solar parameters.

\section{Calculation Results} \label{sec3}

\subsection{Lithium Plateaus Predicted by Rotating Models}

\begin{figure*}[ht!]
\includegraphics[angle=0, scale=0.45 ]{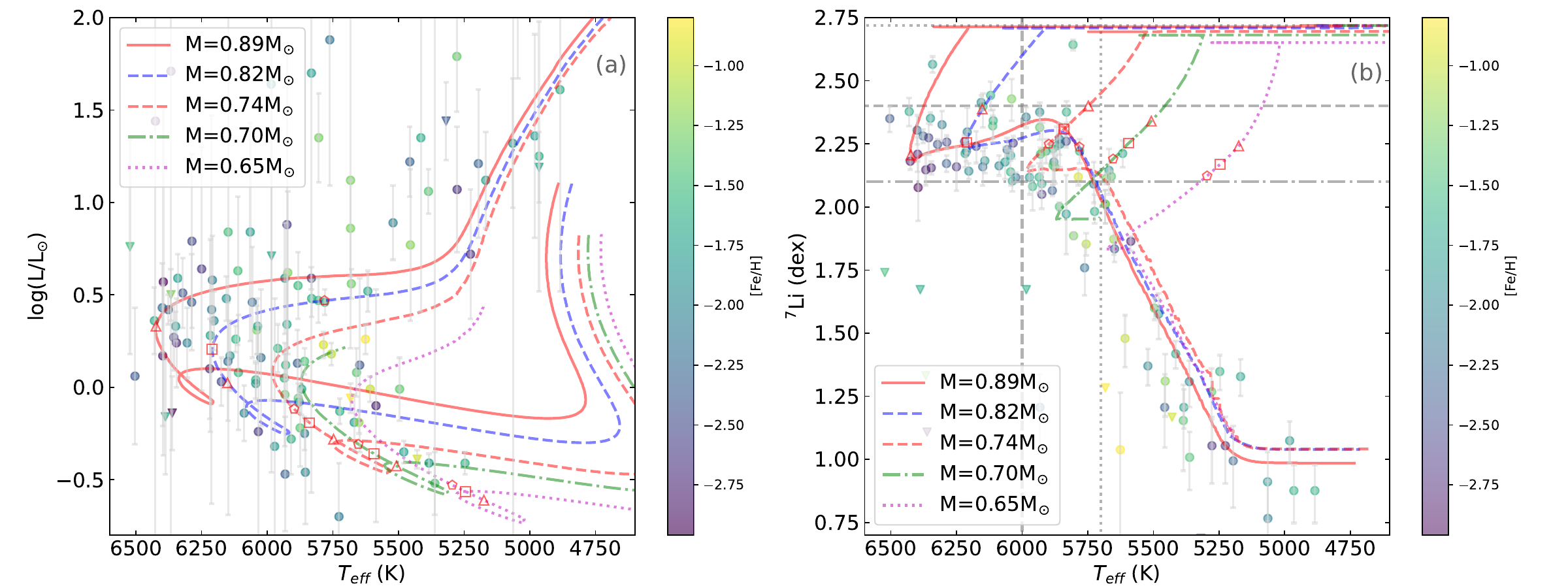}
\caption{(a) Hertzsprung-Russell diagram of rotating models with initial metallicity [Fe/H]$_{0} = -1.0$.
(b) Surface lithium abundance as a function of effective temperature (\teff{}) for different models.
The ages of 8, 11, and 13 Gyr are indicated along the tracks by open red triangles, squares, and pentagons,
respectively. Filled circles denote observed Li abundances, while inverted triangles represent
Li upper limits determined by \citet{charb05}. The metallicities of the observed stars are indicated by
symbol colors. The vertical dashed and dotted lines correspond to \teff{} = 6000 K and 5700 K, respectively.
The horizontal dotted, dashed, and dash-dotted lines indicate Li abundances of 2.72, 2.4, and 2.1 dex, respectively.
\label{fig1}}
\end{figure*}

The [Fe/H] values of the sample studied by \citet{spit82} lie mostly in the range $-0.5$ to $-2.5$.
Assuming solar-scaled abundance mixtures, we first computed the evolution of RMs with an initial
metallicity of [Fe/H]$_{0} = -1.0$ and $M < 1.0$ \dsm{}. Figure \ref{fig1} presents
the Hertzsprung–Russell diagram and surface Li abundances as a function of \teff{} for
these models. Ages of 8, 11, and 13 Gyr are marked along the tracks by open red triangles,
squares, and pentagons, respectively.

Panel (b) of Figure \ref{fig1} shows that the Li abundances of MS and SGB stars with ages between 8 and
13 Gyr and effective temperatures in the range 5700 K $<$ \teff{} $\lesssim$ 6400 K mostly fall between
2.1 and 2.4 dex, in good agreement with the Spite plateau. As relatively massive stars evolve from the
main-sequence turn-off (MSTO)---defined as the point where the effective temperature of an MS star reaches
its maximum---to the middle of the SGB (around 5850 K), their surface Li abundances remain nearly constant.
For example, in the $M = 0.89$ \dsm{} model, whose effective temperature and age at the MSTO are approximately
6400 K and 8 Gyr, respectively, the Li abundance increases slightly from about 2.2 dex to 2.3 dex from the MSTO
to the middle of the SGB, peaking at 2.35 dex around 5900 K before declining steeply until \teff{} $\approx$
5200 K (the base of the red giant branch, RGB). In the $M = 0.74$ \dsm{} model, the Li abundance is 2.31 dex
at 11 Gyr and 2.25 dex at 13 Gyr (see Figure \ref{fig1}). These values are consistent with the mean Li abundance
of $2.224 \pm 0.075$ dex for the Spite plateau reported by \citet{charb05} and $2.24 \pm 0.05$ dex for five TO
stars reported by \citet{korn07}. The Li abundances of the $M = 0.65$ \dsm{} model with ages between 8 and 13 Gyr
also lie within 2.1--2.4 dex, although their effective temperatures and luminosities are comparatively low.

The Li abundance of the $M = 0.74$ \dsm{} model is higher than that of the $M = 0.82$ \dsm{} model at
an age of 11 Gyr (see Table \ref{tab1}). Moreover, stars near the middle of the SGB show higher
Li abundances than MSTO stars (see panel (b) of Figure \ref{fig1} or the $M = 0.82$ \dsm{} models in
Table \ref{tab1}). Thus, the Li abundances of RMs with ages between 8 and 13 Gyr and \teff{} between 5900 K and 6400 K form
a plateau with \ali{} ranging from about 2.2 to 2.4 dex, exhibiting a slight negative slope in \ali{} versus \teff{}.
This feature is consistent with the findings of \citet{korn06, korn07} and \citet{lind09} in NGC 6397.

Due to gravitational settling, the metallicity of the $M = 0.82$ \dsm{} model decreases to approximately
$-1.3$ dex at 11 Gyr, while that of the $M = 0.74$ \dsm{} model is $-1.15$ dex at the same age. Thus, models
on the blue side of the Spite plateau exhibit lower metallicities than those on the red side, even when they
share the same initial metallicity. This predicted metallicity trend is consistent with the observations of
\citet{charb05} (see panel (b) of Figure \ref{fig1}).

Moreover, the Li abundances of dwarf stars with \teff{} $\lesssim$ 5700 K (M $\lesssim$ 0.7 \dsm{}) are lower than
those of MS and SGB stars with \teff{} around 5900 K. For instance, the Li abundance of the $M = 0.70$ \dsm{}
model at 13 Gyr is 2.19 dex, compared with 2.25 dex for the $M = 0.74$ \dsm{} model at the same age.
Furthermore, when \teff{} $\lesssim$ 5200 K, RMs predict another Li plateau with \ali{} $\approx$
1.0 dex for RGB stars, consistent with the finding of \citet{mucc22}, who reported a thin Li plateau
among RGB stars with an average Li abundance of 1.09 $\pm$ 0.07 dex.

Overall, the Li-abundance distribution predicted by RMs aligns well with the observational results of \citet{spit82},
\citet{charb05}, and \citet{mucc22}, suggesting that rotation---including the influence of magnetic fields---plays
a crucial role in the evolution of Li abundance. The rotating models share the same Li-depletion mechanisms
(both gravitational settling and rotational mixing, calibrated to the Sun) as the solar models
in \citet{yang25}, indicating that the solar Li depletion and the Spite plateau may be governed
by the same physical processes.

\subsection{The Effects of Rotation on Lithium Depletion}
\begin{figure*}[ht!]
\includegraphics[angle=0, scale=0.5]{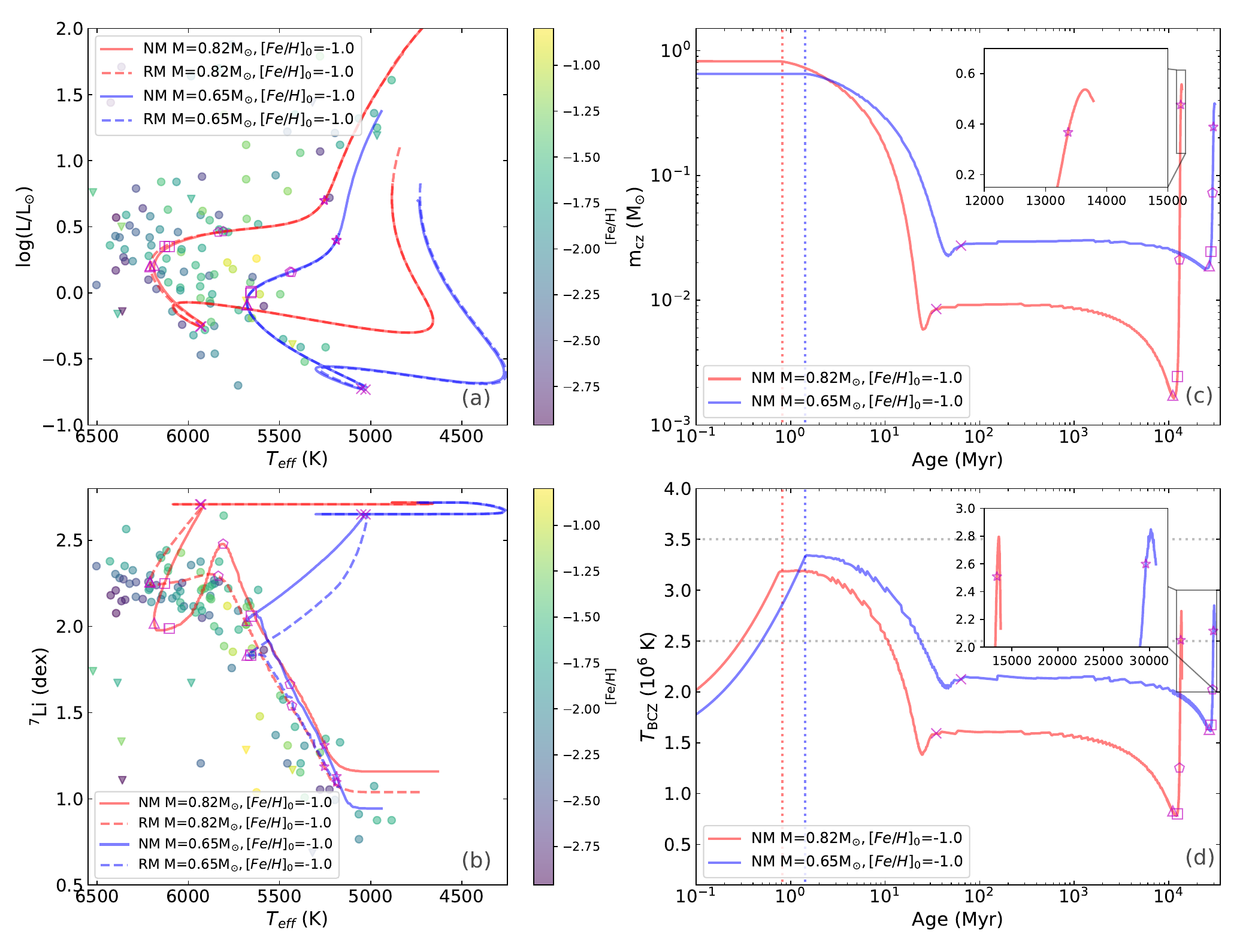}
\caption{(a) Comparison of Hertzsprung-Russell diagrams for rotating and nonrotating models.
Filled circles and inverted triangles represent observational data from \citet{charb05}.
The magenta cross, triangle, square, pentagon, and star along the track mark the ZAMS, MSTO,
terminal-age MS (TAMS), middle of the SGB, and base of the RGB, respectively. The TAMS is
defined as the point where the central hydrogen content drops to $10^{-7}$. These models
are also listed in Tables \ref{tab1} and \ref{tab2}.
(b) Comparison of surface Li abundances between rotating and nonrotating models. Filled circles
denote lithium detections, while inverted triangles indicate Li upper limits \citep{charb05}.
(c, d) Mass of the CZ and temperature at the BCZ of nonrotating models as a function of stellar age.
Vertical dotted lines indicate the end of the fully convective phase.
\label{fig2}}
\end{figure*}

Figure \ref{fig1} clearly shows that models with M $>$ 0.74 \dsm{} follow similar
Li evolution tracks, distinct from those with M $<$ 0.74 \dsm{}. This implyies that the impact of rotation
on Li abundance depends on stellar mass. To understand how the Li plateaus form and the mechanisms driving
them, we compare the evolutions of RMs and nonrotating models (NMs) with $M$ = 0.82 and 0.65 \dsm{}
in Figure \ref{fig2}. Additionally, Li profiles as functions of radius and temperature for different models
are shown in Figure \ref{fig3}.

Rotation affects the structure and evolution of MS stars in two main ways. First, centrifugal
acceleration reduces the effective gravity, lowering both the effective temperature and luminosity.
Second, rotational mixing redistributes chemical elements, leading to a smaller stellar radius
and a higher effective temperature compared with a nonrotating model of the same age \citep{yang13}.
Because the initial rotation rate is low, and both disk-locking and magnetic braking further reduce
rotation, centrifugal effects are negligible. Consequently, rotational mixing dominates, resulting
in slightly higher temperatures and luminosities, although the differences are minor (see panel (a)
of Figure \ref{fig2}). In contrast, rotation exerts a strong influence on Li evolution (see panel
(b) of Figure \ref{fig2}).

\subsubsection{The Effects for Relatively Massive Stars}

To investigate the origin of the differences in Li abundances between RMs and NMs,
panels (c) and (d) of Figure \ref{fig2} show the evolution of the CZ mass (m$_{\mathrm{cz}}$)
and the temperature at the BCZ (\tbc{}), respectively. These panels indicate that from
stellar birth to the RGB stage, the surface remains convective. Moreover, higher stellar
mass corresponds to a shallower CZ and a lower \tbc{}.

At the end of the fully convective stage of the $M = 0.82$ \dsm{} model, \tbc{} is $3.19 \times 10^{6}$ K.
The effective Li-burning temperature for pre-MS stars---that is, the temperature required to
deplete Li within a few $10^{7}$ years---is about $3.5\times10^{6}$ K, which exceeds \tbc{}.
In addition, the timescales of gravitational settling and rotational mixing are much longer
than the duration of the pre-MS stage (about 30 Myr, see panel (c) of Figure \ref{fig2}).
Consequently, the surface Li abundance remains unchanged during this phase.

\begin{figure*}[ht!]
\plotone{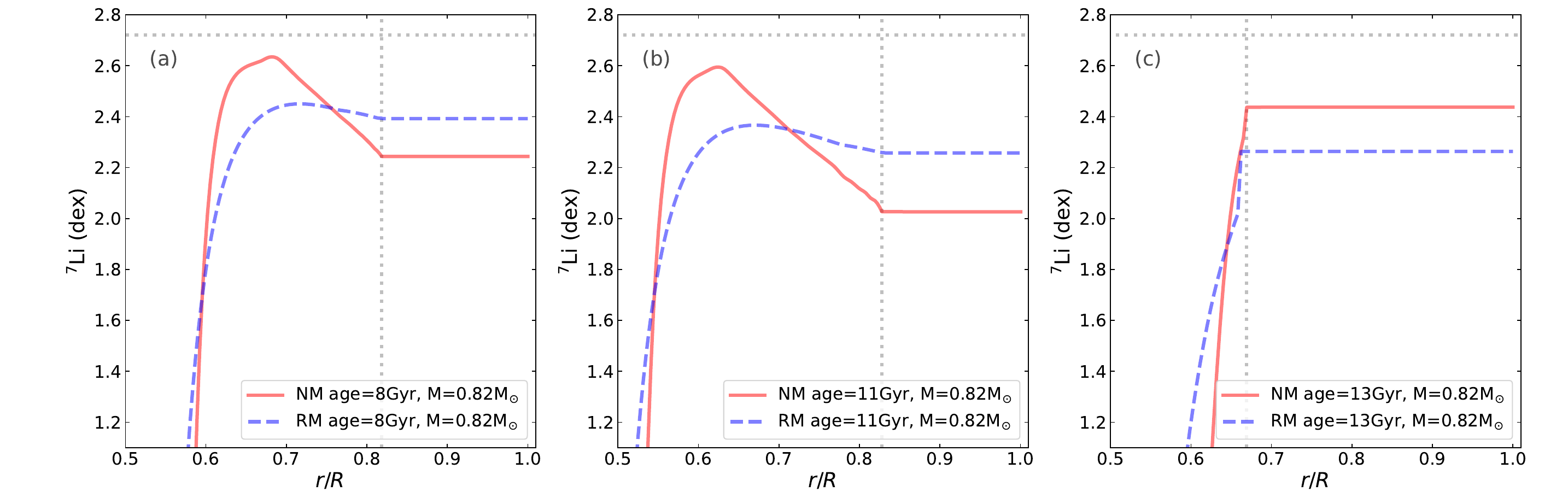}
\plotone{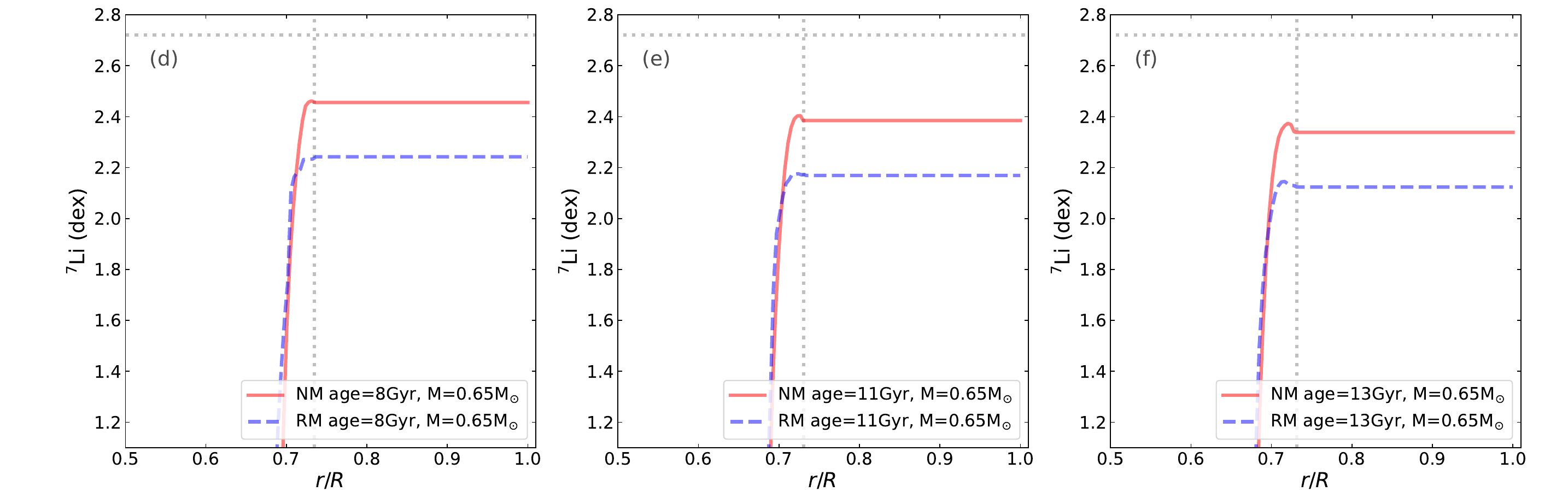}
\plotone{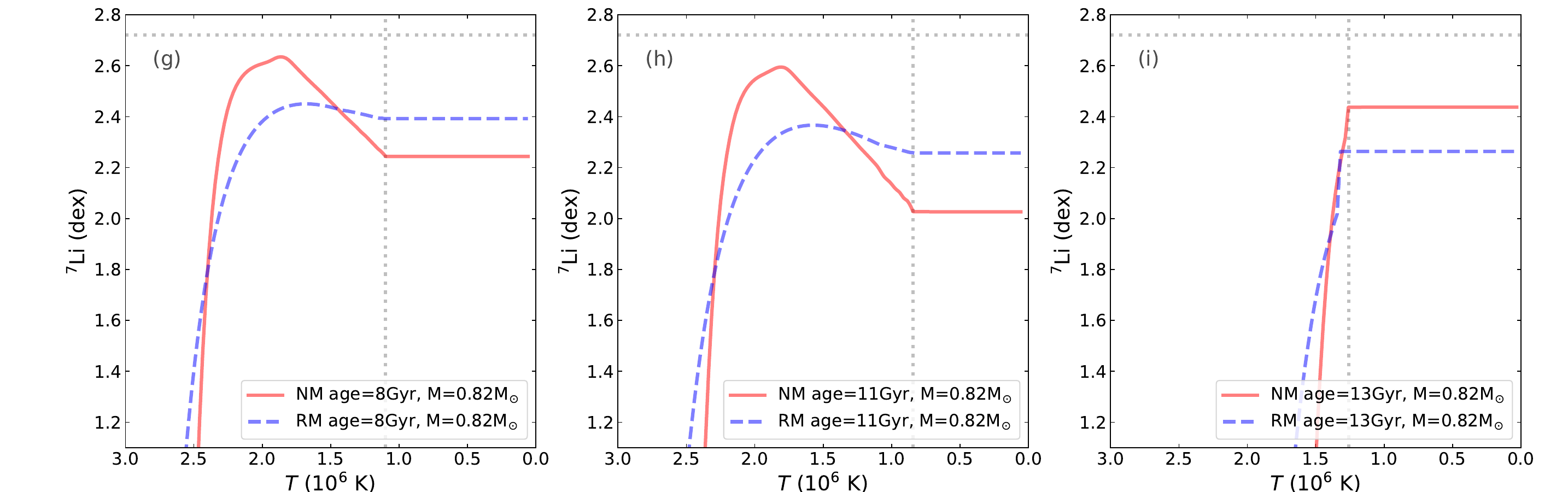}
\plotone{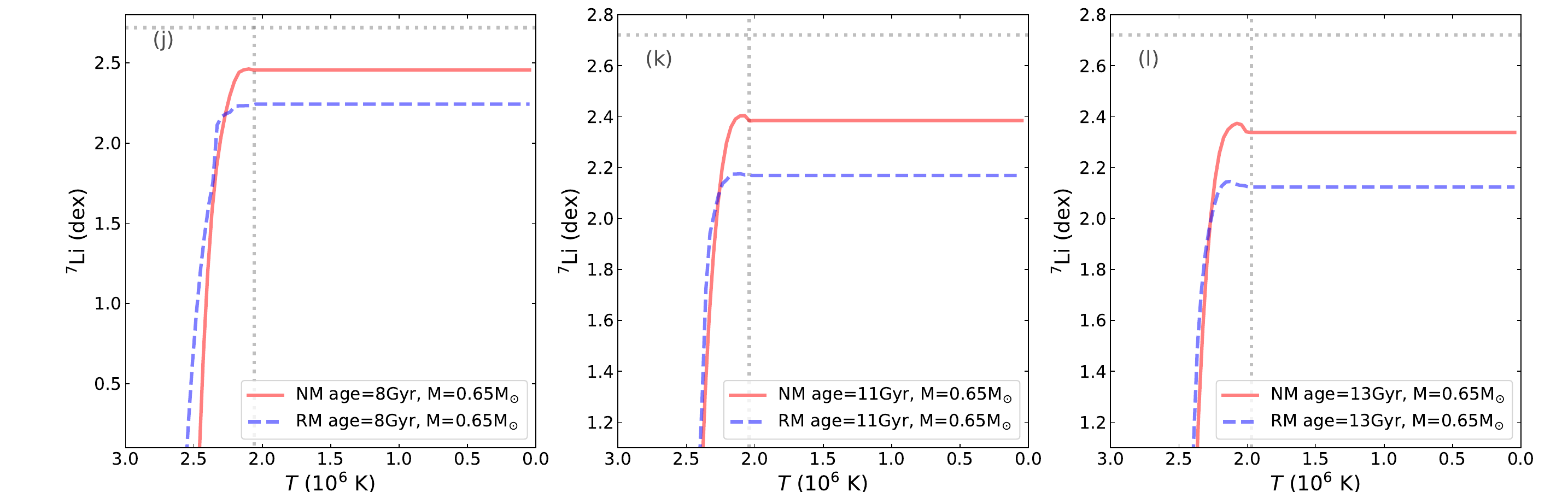}
\caption{(a), (b), (c), (d), (e), (f) Lithium proﬁles as a function of radius for different models.
(g), (h), (i), (j), (k), (l) Lithium proﬁles as a function of temperature for different models.
Horizontal dotted lines indicate the initial Li abundance, while the vertical dotted lines denote
the BCZ of nonrotating models. For M = 0.82 \dsm{}, at ages of 11 and 13 Gyr, the star is located near
the MSTO and the middle of the SGB, respectively.
\label{fig3}}
\end{figure*}

For nonrotating models, the MSTO age of the $M = 0.82$ \dsm{} star is about 11 Gyr. From the ZAMS to
the MSTO, the model develops a progressively shallower CZ, especially near the MSTO. The gravitational
settling timescale of Li in the CZ is proportional to the CZ mass \citep{mich86}; thus, a shallower CZ
corresponds to a stronger settling effect, which becomes most pronounced near the MSTO. Consequently,
the Li abundance decreases from about 2.71 dex at the ZAMS to 2.03 dex at the MSTO as \teff{} or age
increases, due to gravitational settling. During this stage, \tbc{} lies in the range of approximately
$1.5 \times 10^{6} - 1 \times 10^{6}$ K (see panel (d) of Figure \ref{fig2}). The Li burning rate,
$\partial \ln A(\mathrm{Li})/\partial t$, is around $10^{-23}$ s$^{-1}$ at $T = 1.5\times 10^{6}$ K,
indicating that Li burning at the BCZ of the M = 0.82 \dsm{} model is negligible. Thus, Li deposited
below the CZ can be partially preserved. As a result, gravitational settling produces a steeply negative
Li-abundance gradient below the CZ in the NMs with \teff{} $\gtrsim$ 6000 K (see panels of (a) and (b),
or (g) and (h), of Figure \ref{fig3}).

From the MSTO to the middle of the SGB (\teff{} $\approx$ 5800 K and age $\approx$ 13 Gyr for $M = 0.82$ \dsm{}),
the CZ deepens rapidly (see panel (c) of Figure \ref{fig2}). As the CZ deepens, heavy-element settling is weakened.
Moreover, \tbc{} remains too low (see panel (d) of Figure \ref{fig2} or Tables \ref{tab1} and \ref{tab2})
to deplete Li efficiently. The deepening CZ dredges up Li preserved beneath it, leading to a significant increase
in Li abundance as \teff{} decreases, with a pronounced minimum near the MSTO. However, this increase does not
restore the surface Li abundance to its initial value, since part of Li has already settled into deeper layers
and has been destroyed. These results clearly demonstrate that nonrotating evolutionary models cannot account
for the Spite plateau.

By the middle of the SGB, Li below the BCZ is almost completely depleted through diffusion, settling,
and burning, leaving the Li abundance beneath the CZ markedly lower than that within the CZ (see panels (c)
and (i) of Figure \ref{fig3}). Further deepening of the CZ cannot increase \ali{}, and the Li abundance
within the CZ reaches its maximum. From the middle of the SGB to near the base of the RGB, the deepening CZ
both transports Li into hotter regions---where it is rapidly destroyed---and dilutes the Li abundance
in the CZ. The dilution effect dominates, producing a pronounced decrease in Li abundance with decreasing
\teff{}. However, the predicted \ali{} is higher than the observed ones given by \citet{charb05}.

Near the base of the RGB, the rapid expansion of the star halts the rise of \tbc{}, which reaches
a maximum before declining (see panel (d) of Figure \ref{fig2}). Although \tbc{} exceeds $2.5\times10^{6}$ K,
it is still insufficient to deplete Li efficiently on timescales of a few hundred Myr. Gravitational
settling is likewise negligible because the large CZ mass results in an extremely long settling timescale.
Once \tbc{} begins to decrease, Li depletion effectively ceases, leaving the Li abundance nearly
constant for \teff{} $\lesssim$ 5200 K. Nonrotating models predict \ali{} = 1.16 dex for RGB stars,
which is higher than the observed ones reported by \citet{charb05} (see panel (b) of Figure \ref{fig2}).
Thus, Li abundances predicted by NMs are inconsistent with observations, indicating that gravitational
settling and diffusion alone cannot explain the Li problem.

For rotating models, the rotational mixing efficiency calibrated to the Sun is relatively low and insufficient
to fully counteract gravitational settling \citep{yang25}. However, the negative Li-abundance gradient
produced by gravitational settling allows rotational mixing to transport Li from the radiative region
into the CZ. In this way, rotation---including the influence of magnetic fields---mitigates Li depletion
and smooths the Li gradient below the CZ in the $M = 0.82$ \dsm{} model (see panels (a), (b), and
(c) of Figure \ref{fig3}). As a result, the surface \ali{} of rotating MS models is higher than
that of nonrotating MS models (see panel (b) of Figure \ref{fig2}), and the Li abundances of RMs with
ages between 8 and 11 Gyr fall within 2.39--2.26 dex, consistent with the Spite plateau.

In deeper regions (\tbc{} $\gtrsim 1.8 \times 10^{6}$ K), the Li-abundance gradient becomes positive
owing to the rapid increase in the Li-burning rate with temperature and the effects of diffusion.
Consequently, rotational mixing transports Li into hotter layers where it is burned, leading to a
decrease in Li abundance in these zones (see panels (a) and (b), as well as (g) and (h) of Figure
\ref{fig3}). As a result, the Li-abundance gradient below the CZ in RMs is substantially weakened,
yielding an almost flat Li profile. The difference in Li abundance between the CZ and deeper layers
in RMs is therefore much smaller than in NMs. Thus, as RMs evolve from the MSTO (11 Gyr for $M = 0.82$
\dsm{}) to the middle of the SGB ($\sim$13 Gyr), CZ deepening does not lead to a significant Li enhancement;
instead, they maintain an almost constant \ali{} during this phase. For instance, in the $M = 0.82$
\dsm{} model, \ali{} increases only slightly---from 2.26 to 2.3 dex---between the MSTO and the middle
of the SGB, peaking at \teff{} $\simeq$ 5900 K just before a sharp decline. From 8 to 13 Gyr, \ali{}
in the RMs decreases from 2.39 to 2.26 dex and then rises again to 2.3 dex as \teff{} drops markedly,
forming a well-defined Li plateau in the \ali{}--\teff{} plane.

From the middle of the SGB to the RGB (a timescale of $\sim$0.4 Gyr), the CZ mass in the $M = 0.82$
\dsm{} model increases from about 0.021 to 0.367 \dsm{}. The CZ is assumed to be fully mixed.
The impact of CZ deepening on the chemical composition of the CZ far outweighs that of rotational mixing.
Consequently, the evolution of \ali{} in RMs during this phase is governed primarily by CZ deepening.
As in NMs, the Li abundance in RMs declines rapidly during the late SGB stage and then remains nearly
constant until just before the RGB bump. However, because rotational mixing transports Li into hotter
layers where it is more efficiently destroyed, RMs predict \ali{} $\approx$ 1.04 dex on the RGB,
compared with 1.16 dex for NMs. This result agrees more closely with the observations of \citet{charb05}
than the NM prediction.

\subsubsection{The Effects for Lower-mass Stars}

The age of the $M = 0.65$ \dsm{} model at the MSTO is about 27 Gyr, which is much greater than
the age of the universe. Therefore, we focus mainly on its evolution prior to the MSTO. At the
end of the fully convective stage, \tbc{} is approximately $3.34\times 10^{6}$ K, slightly
higher than the $3.19\times 10^{6}$ K of the $M = 0.82$ \dsm{} model. Consequently, the Li abundance
in the $M = 0.65$ \dsm{} model is depleted by about 0.07 dex during the pre-MS stage.

The CZ of the $M = 0.65$ \dsm{} model is significantly deeper than that of the $M = 0.82$ \dsm{} model
(see Figure \ref{fig2}). From the ZAMS to an age of 13 Gyr, \tbc{} exceeds $1.98\times 10^{6}$ K.
On gigayear timescales, Li in regions where $T\gtrsim 2.5 \times 10^{6}$ K is heavily depleted, producing
a positive Li-abundance gradient between the layer with $T \approx 2.5\times 10^{6}$ K and the BCZ
due to the efficient Li burning and diffusion. The negative Li-abundance gradient caused by gravitational
settling is negligible in these models (see Figure \ref{fig3}). Therefore, in the $M = 0.65$ \dsm{} models,
rotational mixing transports Li into hotter regions, enhancing Li depletion. Consequently, RMs with
$M = 0.65$ \dsm{} predict lower Li abundances than their nonrotating counterparts during the MS stage
(see Figures \ref{fig2} and \ref{fig3}).

These behaviors make the $M = 0.65$ \dsm{} models distinctly different from the $M = 0.82$ \dsm{} models,
in which a broad region below the CZ exhibits a markedly negative Li-abundance gradient owing to
the shallower CZ and stronger gravitational settling, while rotational mixing mitigates Li depletion.

For [Fe/H]$_{0}=-1$, the effective temperatures of the $M = 0.74$ \dsm{} models with ages
between 8 and 13 Gyr lie in the range 5700--6000 K. This temperature range represents a turning
point in the effects of rotation on Li abundance. In stars with \teff{} $\lesssim$ 5800 K,
rotational mixing enhances Li depletion, as it does in the Sun \citep{yang25}. A lower stellar
mass corresponds to a higher \tbc{} and a deeper CZ, leading to stronger pre-MS Li depletion
and more efficient Li destruction through rotational mixing. As a result, lower-mass
dwarfs with \teff{} $\lesssim$ 5800 K and [Fe/H]$_{0} =-1.0$ exhibit lower Li abundances.

Compared to the $M = 0.82$ \dsm{} model, the $M = 0.65$ \dsm{} model has almost no lithium available
below the CZ to be dredged up when the star reaches the MSTO. The rapid deepening of the CZ dilutes
Li abundance within it, causing \ali{} to decline primarily with decreasing \teff{} along the SGB.

These results suggest that the Spite plateau arises from the combined effects of variations in CZ
depth, gravitational settling, diffusion, rotational mixing, and magnetic fields. Models that exclude
rotational effects cannot account for the Spite plateau. The impact of rotation---including
magnetic effects---on Li abundance depends on stellar mass or effective temperature. For relatively
massive MS stars with \teff{} $\gtrsim$ 6000 K, such as those with $M = 0.82$ \dsm{}, rotational mixing
partially counteracts gravitational settling and smooths the Li gradient below the CZ during
the MS stage, thereby mitigating Li depletion. As a result, the Li abundances of these stars
remain within a range of about 2.2--2.4 dex for a long time (a few to several gigayears) before
the MSTO. From the MSTO to the middle of the SGB, the Li abundances increase slightly.
These stars constitute the Spite plateau. In contrast, for lower-mass stars with \teff{} $\lesssim$
5800 K, such as those with $M = 0.65$ \dsm{}, rotational mixing enhances Li depletion.

\subsection{The Effects of Metallicity on Lithium Plateaus }

\begin{figure*}[ht!]
\includegraphics[angle=0, scale=0.45 ]{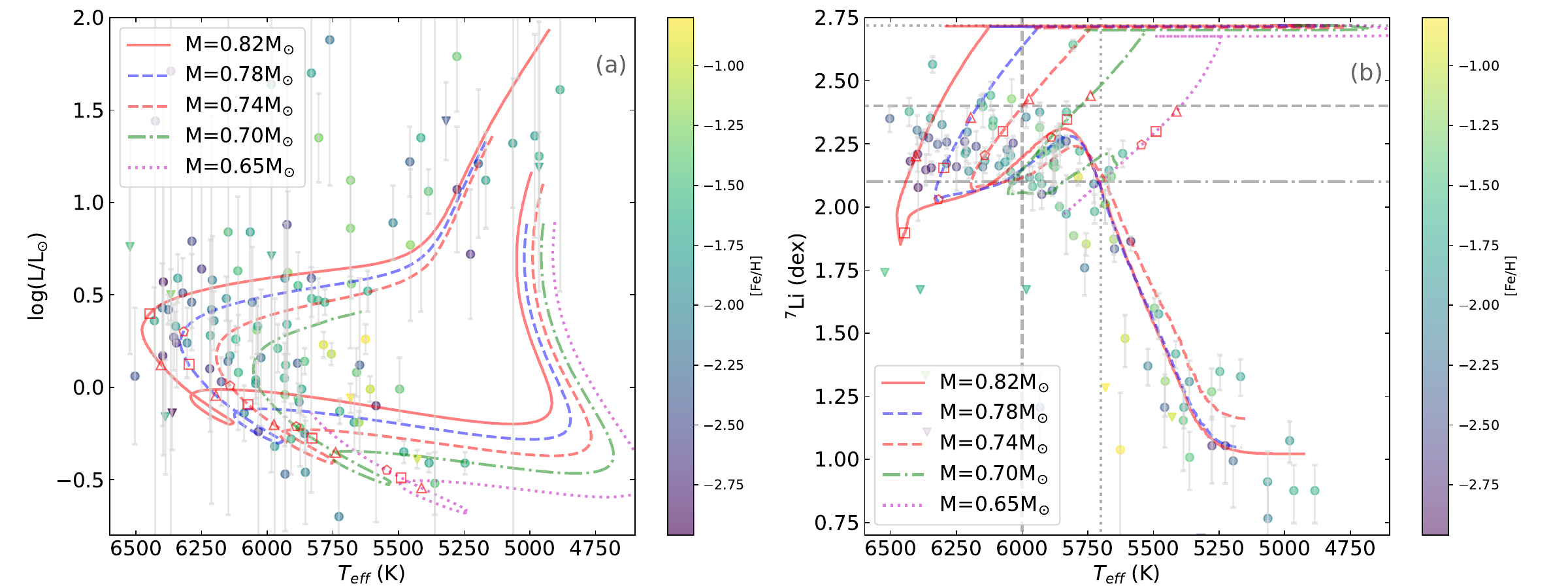}
\includegraphics[angle=0, scale=0.45 ]{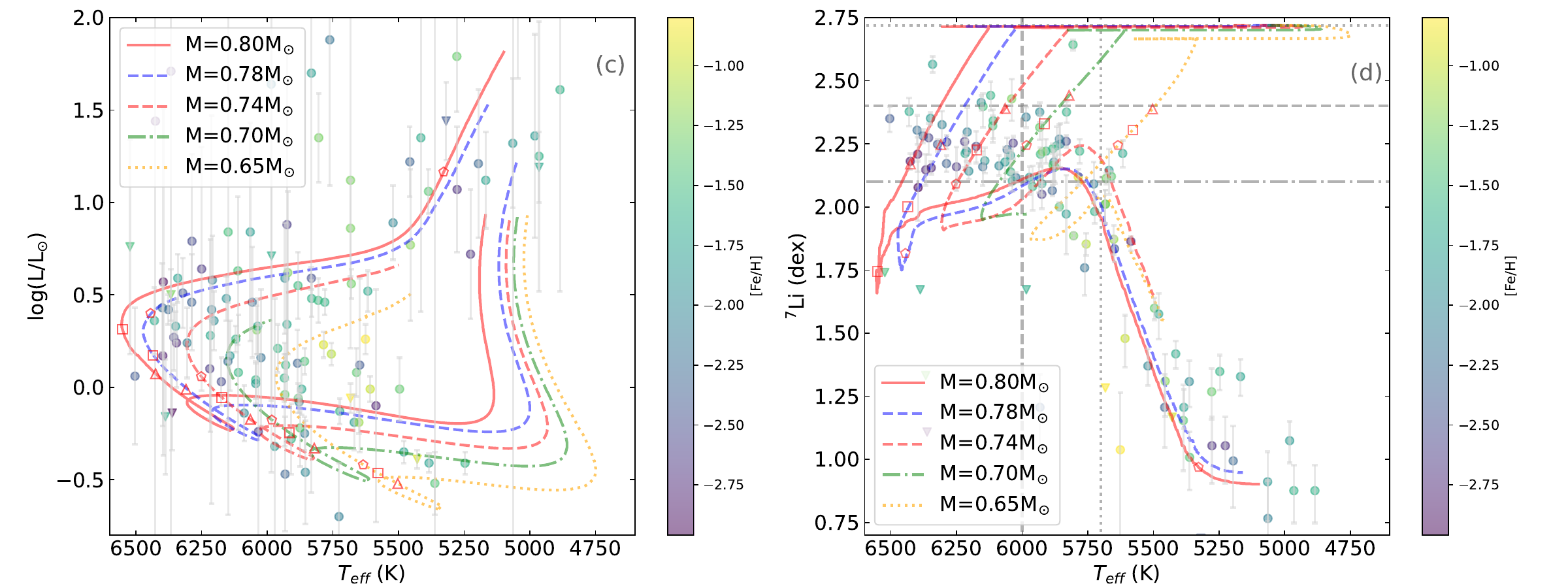}
\caption{Same as Figure \ref{fig1}, but for models with [Fe/H]$_{0} = -1.5$ dex (a, b) and [Fe/H]$_{0} = -2.0$ dex (c, d).
\label{fig4}}
\end{figure*}

Since both stellar mass and metallicity can influence the depth of the CZ, the Li plateau is expected to depend on
metallicity as well. Using solar-scaled mixtures, we computed evolutionary models with initial metallicities
of [Fe/H]$_{0} = -1.5$ dex and [Fe/H]$_{0} = -2.0$ dex. Due to gravitational settling, the metallicity of stars near
the MSTO becomes significantly lower than the initial value. For instance, at an age of 13 Gyr, the metallicity
of the $M = 0.78$ \dsm{} model is approximately $-1.9$ dex for [Fe/H]$_{0} = -1.5$ dex and $-2.6$ dex for
[Fe/H]$_{0} = -2.0$ dex. At the same age, the $M = 0.70$ \dsm{} model has metallicities of about $-1.7$ dex
and $-2.2$ dex for the respective initial values. These results indicate that, because the gravitational settling
timescale is proportional to the CZ mass, stars on the blue side of the Spite plateau are more likely to appear
more metal-poor than those on the red side, and that the gravitational settling effects are more pronounced
in very metal-poor stars.

Figure \ref{fig4} shows the Hertzsprung-Russell diagrams and Li-evolution tracks of these models.
It is evident that Li abundances of RMs with 5800 K $\lesssim$ \teff{} $\lesssim$6400 K and ages
between 8 Gyr and 13 Gyr mostly fall within the range of 2.0--2.4 dex, forming a Li plateau in \ali{}--\teff{}
plane. Moreover, decreasing metallicity has little impact on the Li abundances of MS stars on the red
side of the Spite plateau. For example, at 13 Gyr, the Li abundance of the $M=0.70$ \dsm{} model
is 2.28 dex for [Fe/H]$_{0} = -1.5$ dex and 2.24 dex for [Fe/H]$_{0} = -2.0$ dex. However,
the effect becomes much stronger on the blue side of the Spite plateau. At the MSTO (age $\simeq$ 13 Gyr),
the Li abundance of the $M=0.78$ \dsm{} model is 2.03 dex for [Fe/H]$_{0} = -1.5$ dex but only 1.75 dex
for [Fe/H]$_{0} = -2.0$ dex (see panels (b) and (d) of Figure \ref{fig4}). The $M=0.80$ \dsm{} model
with [Fe/H]$_{0} = -2.0$ dex yields \ali{} $\simeq$ 1.66 dex at the MSTO (age $\simeq$ 11 Gyr, [Fe/H]
$\simeq -2.73$ dex). More massive stars exhibit lower Li abundances.

These results indicate that some very metal-poor stars can fall below the blue side of the Spite plateau,
and confirm that \ali{} depends on both \teff{} and metallicity. This is consistent with the findings
of \citet{deli93}, \citet{norr94}, and \citet{thor94}. On the Spite plateau, \ali{} increases
slightly with decreasing \teff{}, and this dependence becomes more pronounced at lower metallicity
(see Figures \ref{fig1} and \ref{fig4}). In other words, the slope of \ali{} versus \teff{} is somewhat
steeper for very metal-poor stars than for moderately metal-poor ones.

In addition, lower metallicity corresponds to lower Li abundance in TO stars. Hence, theoretical
models predict that stars with lower metal content are more likely to lie near the bottom of
the blue side of the Spite plateau, in agreement with the observations of \citet{charb05} (see panels (b)
and (d) of Figure \ref{fig4}). Furthermore, the Li abundances of very metal-poor stars ([Fe/H] $< -2.0$;
models in panel (d)) are more scattered than those of moderately metal-poor stars ([Fe/H] $> -2.0$; models
in panel (b)). This trend is consistent with the findings of \citet{aspl06} and \citet{sbor10}.

\begin{figure*}[ht!]
\includegraphics[angle=0, scale=0.37 ]{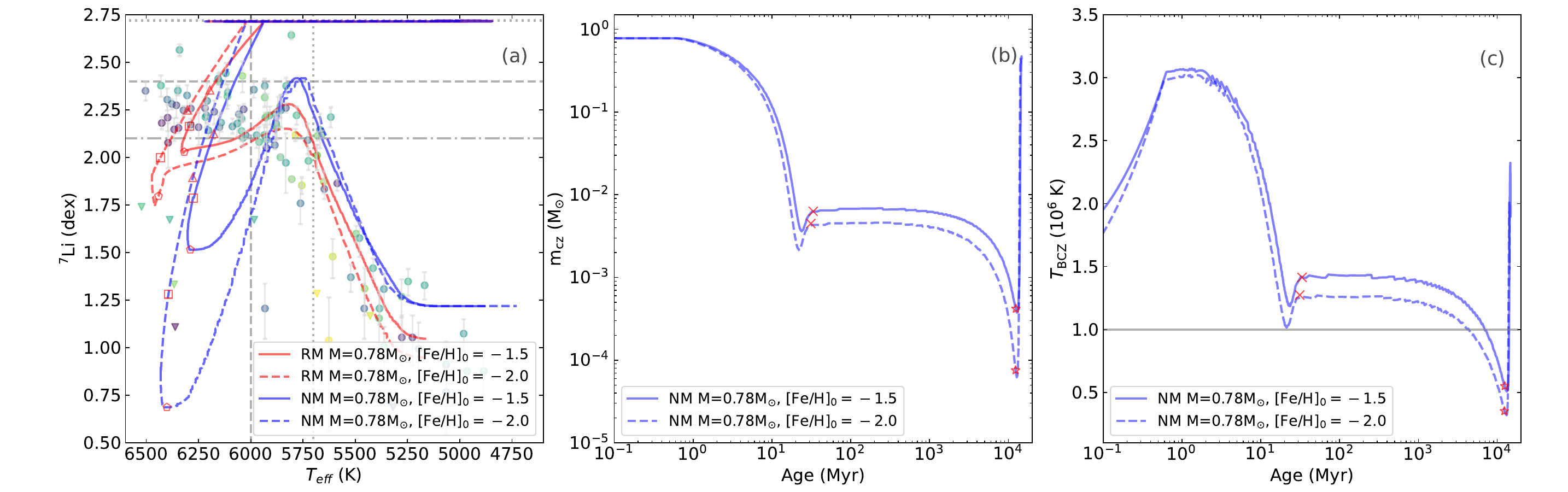}
\caption{(a) Surface lithium abundance as a function of effective temperature for different models. The ages
of 8, 11, and 13 Gyr are indicated along the tracks by open red triangles, squares, and pentagons, respectively.
(b, c) Mass of the CZ and temperature at the BCZ of NMs as a function of stellar age. The cross and star
along the tracks mark the ZAMS and MSTO, respectively.
\label{fig5}}
\end{figure*}

To investigate the origin of the dependence of \ali{} on metallicity and effective temperature,
Figure \ref{fig5} compares the evolutions of \ali{}, m$_{\mathrm{cz}}$, and \tbc{} for models of the same
mass but different metallicities. The results show that the CZ depth decreases with decreasing metallicity;
in other words, the lower the initial metallicity, the shallower the CZ. The CZ depth also decreases rapidly
with increasing age or \teff{} prior to the MSTO, reaching a minimum near the MSTO. The effect is more
pronounced at lower metallicity. For example, at the ZAMS, the CZ mass of the $M = 0.78$ \dsm{} model is
$6.3\times10^{-3}$ \dsm{} for [Fe/H]$_{0} = -1.5$ dex and $4.5\times10^{-3}$ \dsm{} for [Fe/H]$_{0} = -2.0$
dex. At the MSTO, the CZ mass is approximately $4\times10^{-4}$ \dsm{} for [Fe/H]$_{0} = -1.5$ dex and
$8\times10^{-5}$ \dsm{} for [Fe/H]$_{0} = -2.0$ dex, the latter being significantly lower than the former.
Since the gravitational settling timescale of heavy elements in the CZ is proportional to the CZ mass,
settling is more effective in the [Fe/H]$_{0} = -2.0$ dex models than in the [Fe/H]$_{0} = -1.5$ dex models.
Consequently, the [Fe/H]$_{0} = -2.0$ dex models show lower Li abundances, particularly at the MSTO.

The temperature below the CZ of these models remains sufficiently low---below the Li-burning threshold of
$10^{6}$ K for most of the MS stage (see panel (c) of Figure \ref{fig5})---to preserve deposited
Li in a buffer region beneath the CZ. Therefore, as NMs evolve from the MSTO to the middle of the SGB
(\teff{} $\approx 5800$ K), their Li abundances increase rapidly due to dredge-up associated with CZ deepening.
However, this rapid variation leads to a significant discrepancy with observations (see Figure \ref{fig5}),
as the predicted Li abundances are too low. This further rules out the possibility that NMs can account
for the Spite plateau.

In rotating models, rotational mixing partially counteracts gravitational settling, substantially reducing Li depletion
in the CZ. As a result, the CZ Li abundances of RMs between 8 and 13 Gyr are much higher than those of NMs,
generally falling within the range of 1.8--2.4 dex (see Figure \ref{fig5}). This indicates that rotation plays
a crucial role in the formation of the Spite plateau.

\begin{figure*}[ht!]
\includegraphics[angle=0, scale=0.37 ]{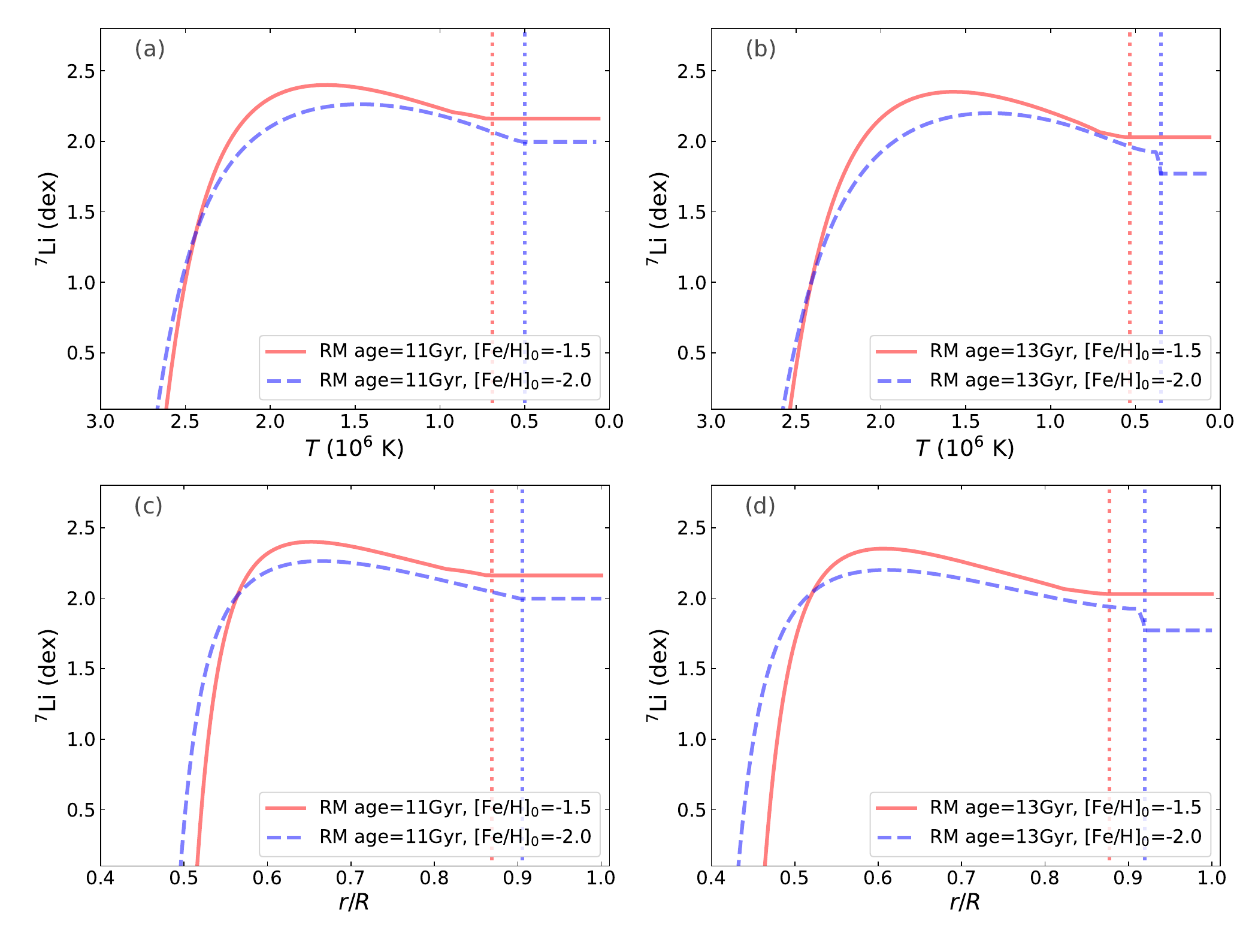}
\caption{(a, b) Lithium abundance profile as a function of temperature for the models with M = 0.78 \dsm{}.
(b, c) Lithium abundance profile as a function of radius for the models with M = 0.78 \dsm{}. The vertical
dotted lines denote the BCZ of the models.
\label{fig6}}
\end{figure*}

Because \tbc{} is sufficiently low and gravitational settling is efficient, these models develop a negative
Li-abundance gradient over a wide region. The rotational mixing calibrated to the Sun is inadequate to eliminate
this gradient. As a result, when these stars evolve to the MSTO, they still maintain a relatively steep negative
Li-abundance gradient in a broad region (from about 0.6 R to 0.9 R) below the CZ (see Figure \ref{fig6}),
compared with the models of [Fe/H]$_{0} = -1.0$ dex (see Figure \ref{fig3}). As the stars
evolve from the MSTO to the middle of the SGB, Li preserved below the CZ is dredged up. Consequently, the Li
abundances of RMs with [Fe/H]$_{0} = -1.5$ dex and $-2.0$ dex increase slightly with decreasing \teff{}.
However, this increase is more pronounced than that in RMs with [Fe/H]$_{0} = -1.0$ dex. In other words,
the dependence of \ali{} on \teff{} becomes stronger as metallicity decreases.

\begin{figure*}[ht!]
\includegraphics[angle=0, scale=0.45]{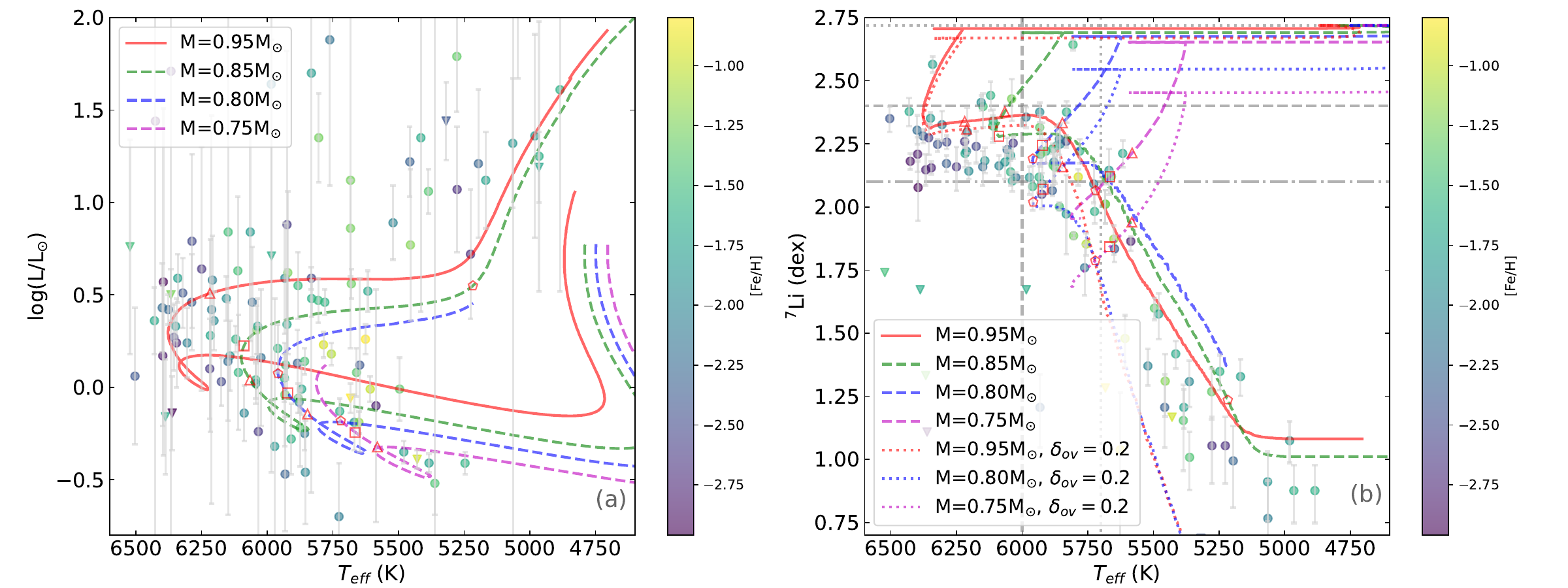}
\caption{Same as Figure \ref{fig1}, but for models with [Fe/H]$_{0} = - 0.7$ dex.
\label{fig7}}
\end{figure*}

The sample in \citet{spit82} and \citet{charb05} include stars with [Fe/H] $> -1.0$ dex. To further
test the dependence of \ali{} on both \teff{} and metallicity, we calculated the evolutions of models
with [Fe/H]$_{0} = -0.7$ dex. The results are shown in Figure \ref{fig7}. Similar to the models
with [Fe/H]$_{0} = -1.0$ dex, the Li abundances of models with \teff{} between about 6400 and 5900 K and
ages between 8 and 13 Gyr mainly fall within 2.23--2.4 dex, forming a tight Li plateau with almost no
dependence on \teff{}. For example, at 8 Gyr, the Li abundance is 2.34 dex for $M = 0.95$ \dsm{},
2.38 dex for $M = 0.85$ \dsm{}, and 2.33 dex for $M = 0.80$ \dsm{}, with a maximum at \teff{} $\simeq$
6050 K. Moreover, as these models evolve from the MSTO to the middle of the SGB, their Li abundances
remain nearly constant (see panel (b) of Figure \ref{fig7}) and are higher than
those of SGB counterparts with [Fe/H]$_{0} < -1.0$ dex (see Figure \ref{fig4}).

These results indicate that the Li abundance of the Spite plateau increases slightly with
metallicity, while the dependence of \ali{} on \teff{} weakens as metallicity increases.
Moreover, they suggest that stars with relatively low metallicity are more likely to appear
near the lower edge of the blue side of the Spite plateau.

For dwarf stars with \teff{} $\lesssim 6000$ K, Li abundances decrease with decreasing mass. For instance,
between 8 and 11 Gyr, the $M \lesssim 0.80$ \dsm{} models predict lower Li abundances than the $M = 0.85$
\dsm{} model (see Figure \ref{fig7}). For these stars, \tbc{} is high enough to deplete Li during the
pre-MS stage. The lower the mass, the deeper the CZ, and the higher \tbc{}, leading to stronger pre-MS
Li depletion that increases with decreasing mass. Additionally, as in the $M = 0.65$ \dsm{} model with
[Fe/H]$_{0} = -1.0$ dex, rotational mixing further enhances Li depletion in these dwarfs. Consequently,
the Li abundances of these stars decrease with decreasing mass or \teff{} (see Figure \ref{fig7}).
The higher the metallicity, the more pronounced this trend becomes due to the deeper CZ.

\subsection{The Effects of Rotation on Other Element Abundances}

\begin{figure*}[ht!]
\plotone{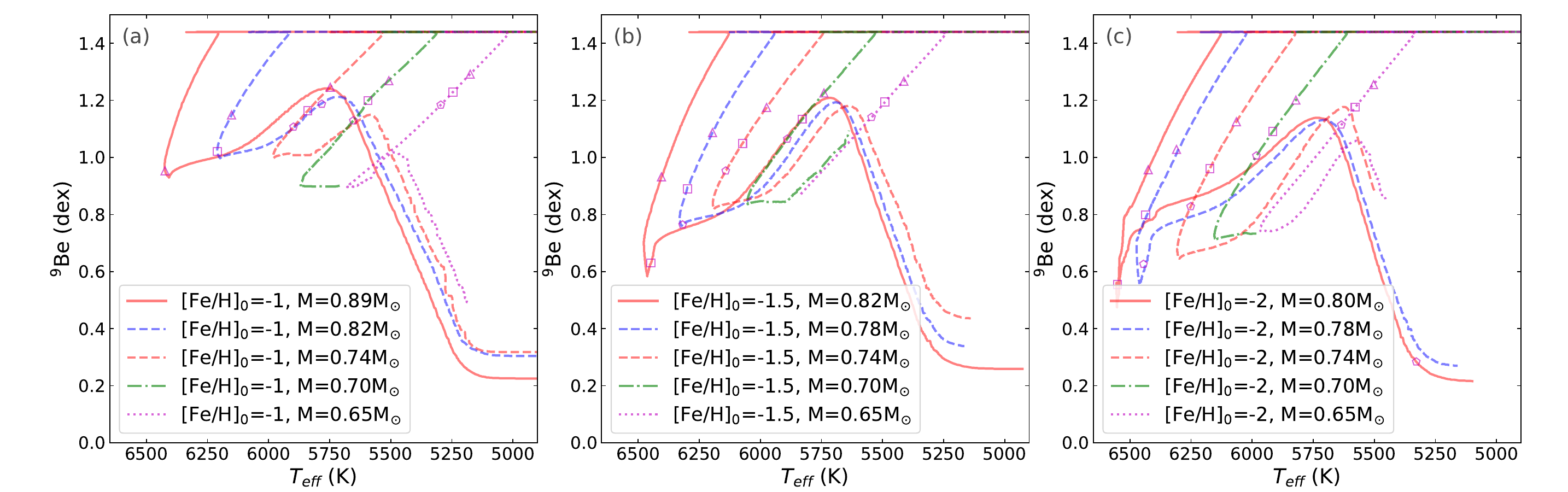}
\plotone{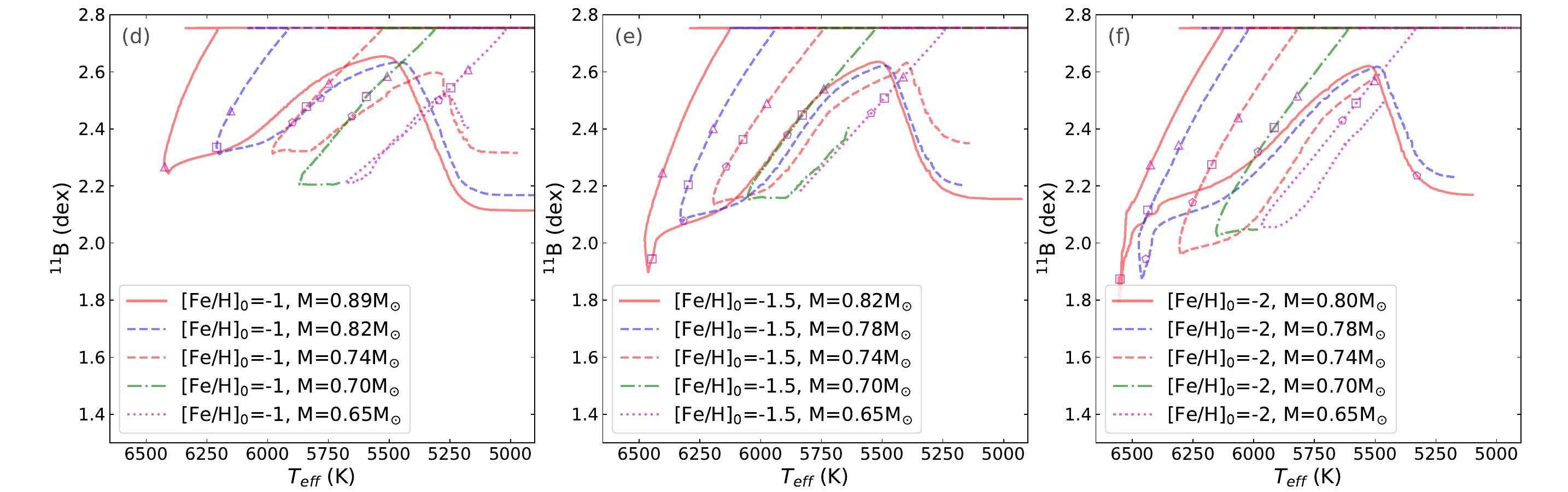}
\caption{(a), (b), (c) Surface $^{9}$Be abundance as a function of \teff{} for different models.
(d), (e), (f) Surface $^{11}$B abundance as a function of \teff{} for different models.
\label{fig8}}
\end{figure*}

\begin{figure*}[ht!]
\plotone{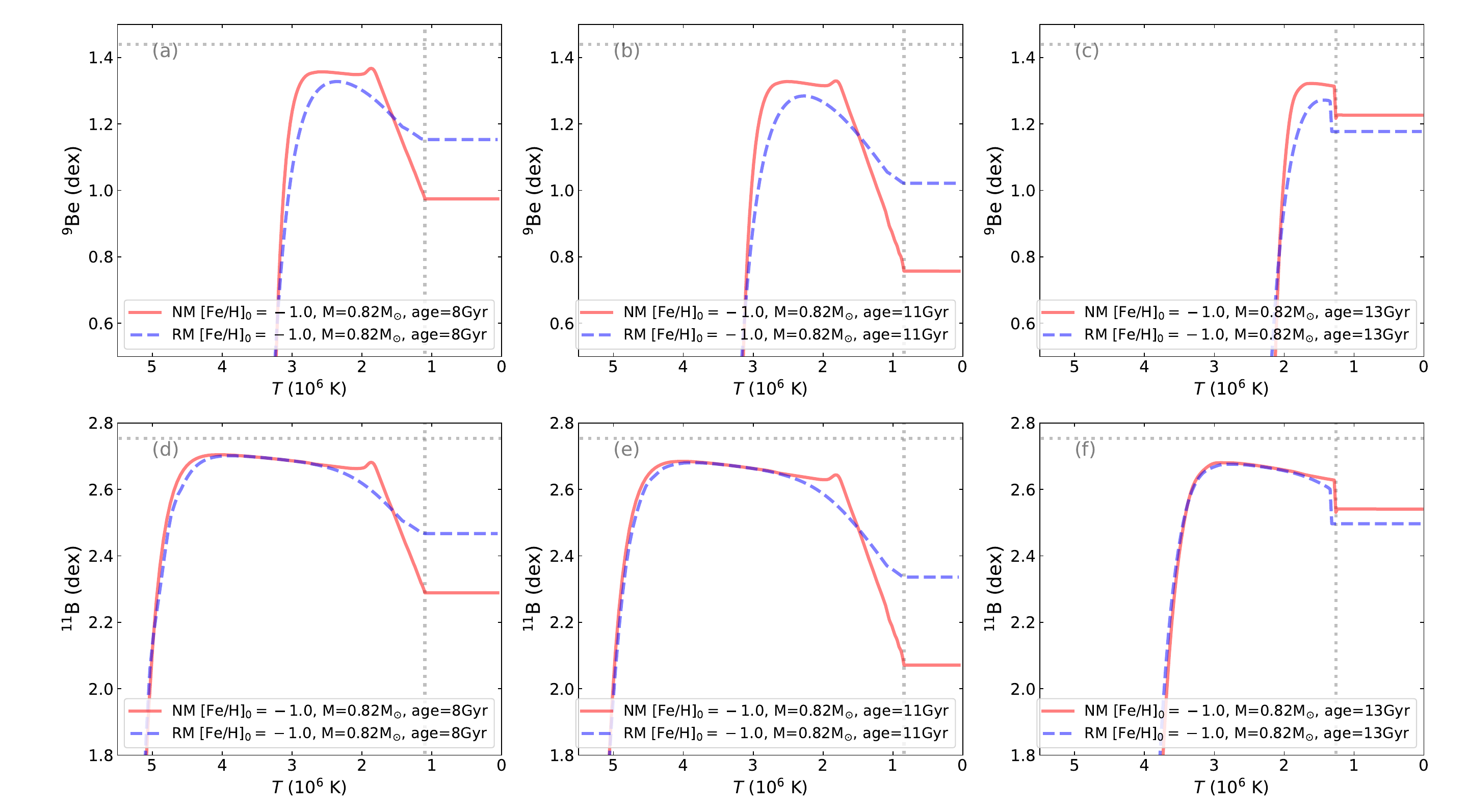}
\caption{(a), (b), (c) $^{9}$Be profiles as a function of temperature for different models.
(d), (e), (f) $^{11}$B profiles as a function of temperature for different models. Horizontal
dotted lines indicate the initial abundance, while the vertical dotted lines denote the base of the surface
CZ of non-rotating models.
\label{fig9}}
\end{figure*}

Beryllium and boron are also fragile elements. The initial abundances of Be and B in halo stars
are unknown; therefore, we assume that they have the same Be and B abundances as the Sun. Figure \ref{fig8}
shows that their behaviors resemble that of Li, in that they also form plateaus. However,
the Be and B plateaus exhibit more pronounced negative slopes
in $A(\mathrm{Be})$ or $A(\mathrm{B})$ versus \teff{}, and their dilutions set in at lower effective
temperatures than that of Li. For instance, in the $M = 0.82$ \dsm{} model with [Fe/H]$_{0} = - 1.0$ dex,
Be dilution begins at \teff{} $\approx 5700$ K, while B dilution occurs at \teff{} $\approx 5450$ K.
This difference arises because Be and B burn at higher temperatures ($3.5\times10^{6}$ K for Be and
$5\times10^{6}$ K for B) than Li. Their higher burning thresholds allow them to survive in hotter regions
where Li is already depleted. For example, in the $M = 0.82$ \dsm{} models with [Fe/H]$_{0} = - 1.0$ dex,
Li below the CZ is almost entirely destroyed by 13 Gyr (see Figure \ref{fig3}), yet significant amounts of
Be and B remain at the BCZ, with B surviving in deeper layers than Be (see Figure \ref{fig9}).
Consequently, Be and B dilution due to CZ deepening occurs later---or at lower \teff{}---than that of Li.
In general, the higher the burning temperature of an element, the later its dilution occurs.

In the $M = 0.65$ \dsm{} model with [Fe/H]$_{0} = -1.0$ dex, \tbc{} is significantly lower than
the effective Be- and B-burning temperatures during the MS stage. The combination of high burning
thresholds and gravitational settling produces negative Be- and B-abundance gradients below the CZ,
unlike Li. Rotational mixing alleviates Be and B settling but cannot erase the negative gradients in
these models, as is also the case for the $M = 0.82$ \dsm{} models and the Sun \citep{yang25}.
In other words, in both the $M = 0.65$ \dsm{} and $M = 0.82$ \dsm{} models, rotational mixing
mitigates the depletions of Be and B. In addition, lower stellar mass corresponds to a deeper CZ and
weaker settling. Consequently, the Be and B abundances of lower-mass stars at ages between
8 and 13 Gyr exceed those of more massive stars. Furthermore, the Be and B abundances increase
markedly as the CZ deepens after the MSTO until dilution begins (see Figure \ref{fig8}).
Accordingly, Be abundance exhibits a steeper negative slope in $A$(Be) versus \teff{} between
about 6400 K and 5700 K, while B abundance shows a steeper negative slope in $A$(B) versus
\teff{} between about 6400 K and 5500 K than does Li abundance.

\begin{figure*}[ht!]
\plotone{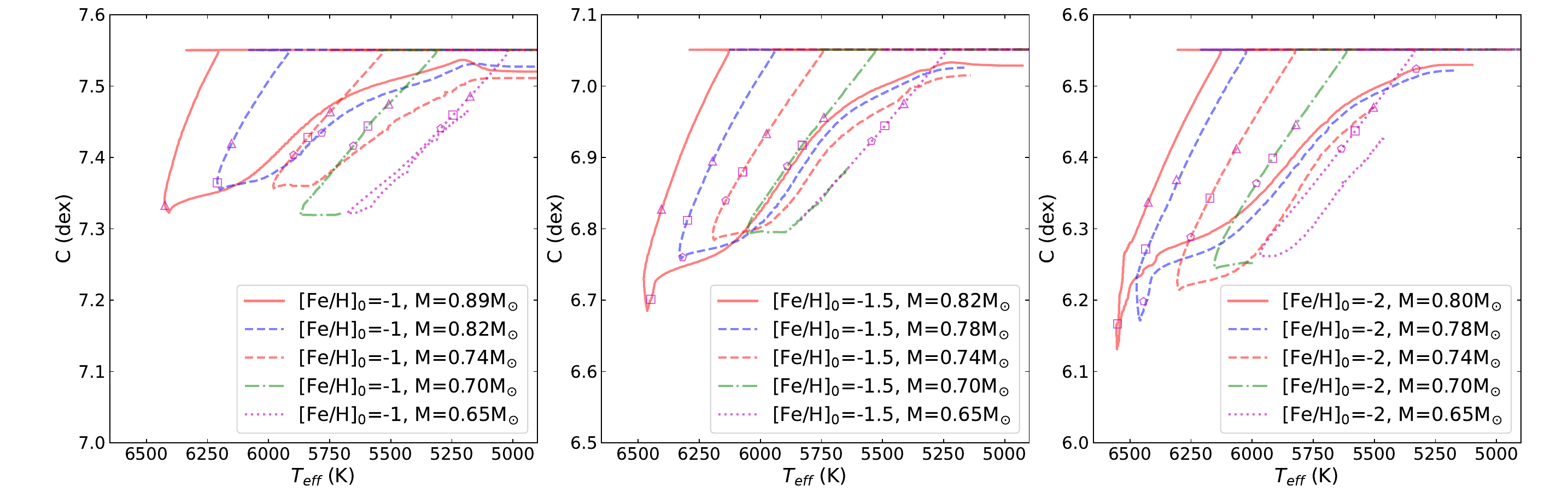}
\plotone{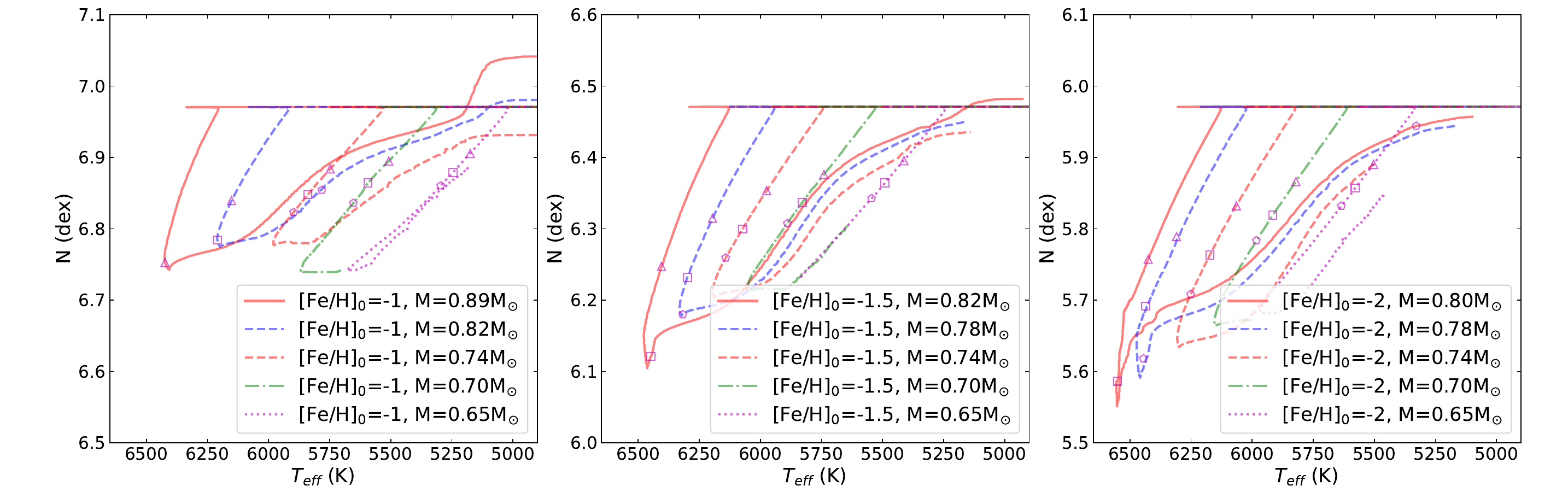}
\plotone{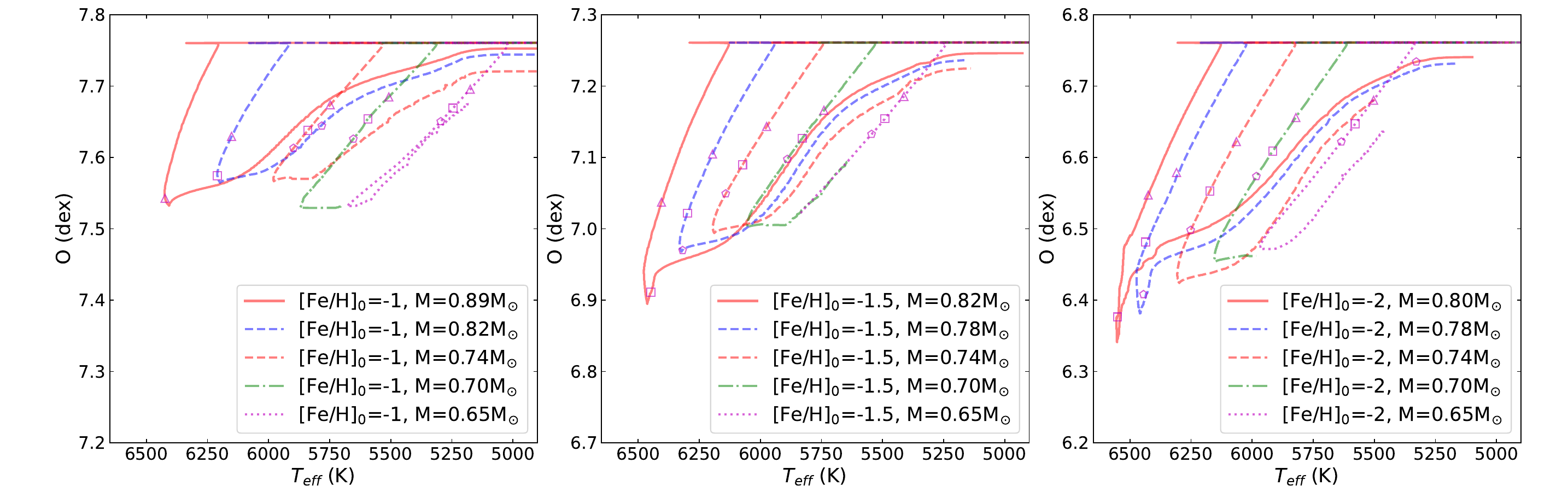}
\caption{Top panels: Carbon abundance as a function of \teff{} for different models.
Middle panels: Nitrogen abundance as a function of \teff{} for different models.
Bottom panels: Oxygen abundance as a function of \teff{} for different models.
\label{fig10}}
\end{figure*}

Elements heavier than B, such as carbon (C), nitrogen (N), and oxygen (O), cannot be burned in these low-mass stars.
Instead, C, N, and O deposited from the CZ are preserved in a buffer region below it. When a star evolves from
the MSTO to the RGB, these elements are dredged up, producing an increase in heavy-element abundance with
decreasing \teff{} during the SGB and RGB phases (see Figure \ref{fig10}). For example, from the MSTO to \teff{}
= 5600 K, the $A$(O) of the $M = 0.78$ \dsm{} model with [Fe/H]$_{0} = - 2.0$ dex increases from 6.38 to 6.68 dex.
These elements cannot be diluted during the SGB and RGB stages. Because Li and Be are strongly diluted during
the late SGB, their abundances are anti-correlated with those of heavier elements in this stage.

\subsection{The Effects of Rotation Rate and $\alpha$-element Enhancement}

\citet{pran18} and \citet{roma19} investigated the impact of rotating massive stars on chemical evolution.
They found that the yields of rotating massive stars have a dramatic effect on the predicted evolution of
s-process elements, but only a minor impact on the evolution of $\alpha$-elements. In this work, we focus
primarily on the evolution of low-mass stars during the hydrogen-burning stage.

\citet{gall15} showed that about $50\%$ of PMS stars with $M \simeq 0.8$ \dsm{} have anglular velocities
lower than approximately $4\Omega_{\odot}$ at ages of 1--2 Myr. Accordingly, we computed evolutionary models
with an initial angular velocity of $\Omega_{i} = 1.1 \times 10^{-5}$ rad s$^{-1}$, corresponding to
$\approx 4.1 \Omega_{\odot}$ or an initial velocity of about 40 km s$^{-1}$). Our results are only weakly
affected by variations in $\Omega_{i}$.

This is because stars exhibiting the Spite plateau possess deep CZs, leading to substantial
angular momentum loss during the early MS phase (before $\sim$ 1 Gyr, see Figure 5 of \citealt{yang25}
or Figure 9 of \citet{chab95}). The larger the initial rotation rate, the higher the rate of angular
momentum loss. Consequently, these stars predominantly appear as slow rotators during the MS stage,
excluding binary systems or merged stars. On the other hand, the Li depletion rate reaches a maximum
near the MSTO; that is, Li depletion mainly occurs during the middle of the MS stage, depending on
stellar mass and metallicity. As a result, variations in $\Omega_{i}$ have only a minor effect on Li depletion.

Metal-poor halo stars are known to exhibit $\alpha$-element enhancement. For very metal-poor stars, \cite{cayr04}
and \citet{spit05} derived abundance ratios of [C/Fe] = 0.18, [O/Fe] = 0.47, [Mg/Fe] = 0.27, [Si/Fe] = 0.37, [S/Fe] = 0.35,
[Ar/Fe] = 0.35, [Ca/Fe] = 0.33, and [Ti/Fe] = 0.23. The abundances of other elements are assumed to scale with
the solar mixture of \citet{magg22}. As a result of these enhancements, the total metallicities corresponding
to [Fe/H] = $-1$, $-1.5$, $-2$ are $Z = 3.456 \times 10^{-3}$, $1.103 \times 10^{-3}$, $3.499 \times 10^{-4}$,
respectively. These values are slightly higher than $3.236 \times 10^{-3}$ for [Fe/H] = $-1$ and $3.236 \times 10^{-4}$
for [Fe/H] = $-2$ reported by \citet{pran18}, owing to our adoption of a different solar mixture.

\begin{figure*}[ht!]
\includegraphics[angle=0, scale=0.35]{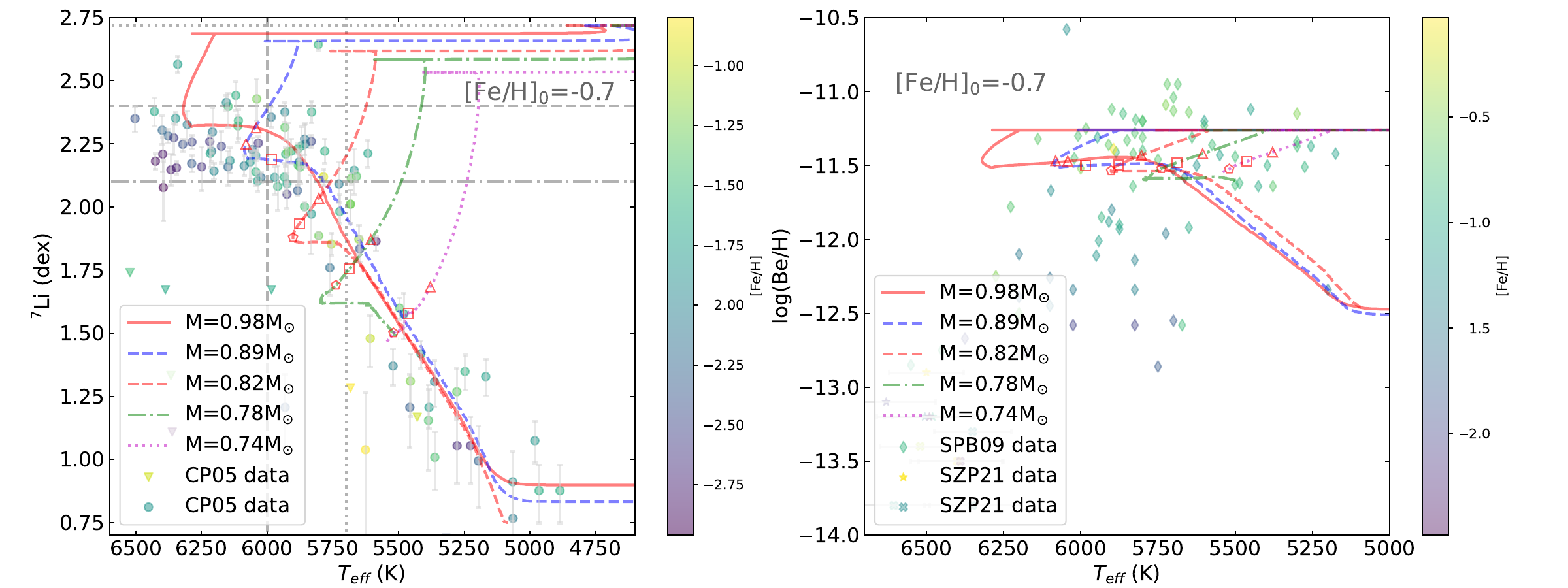}
\includegraphics[angle=0, scale=0.35]{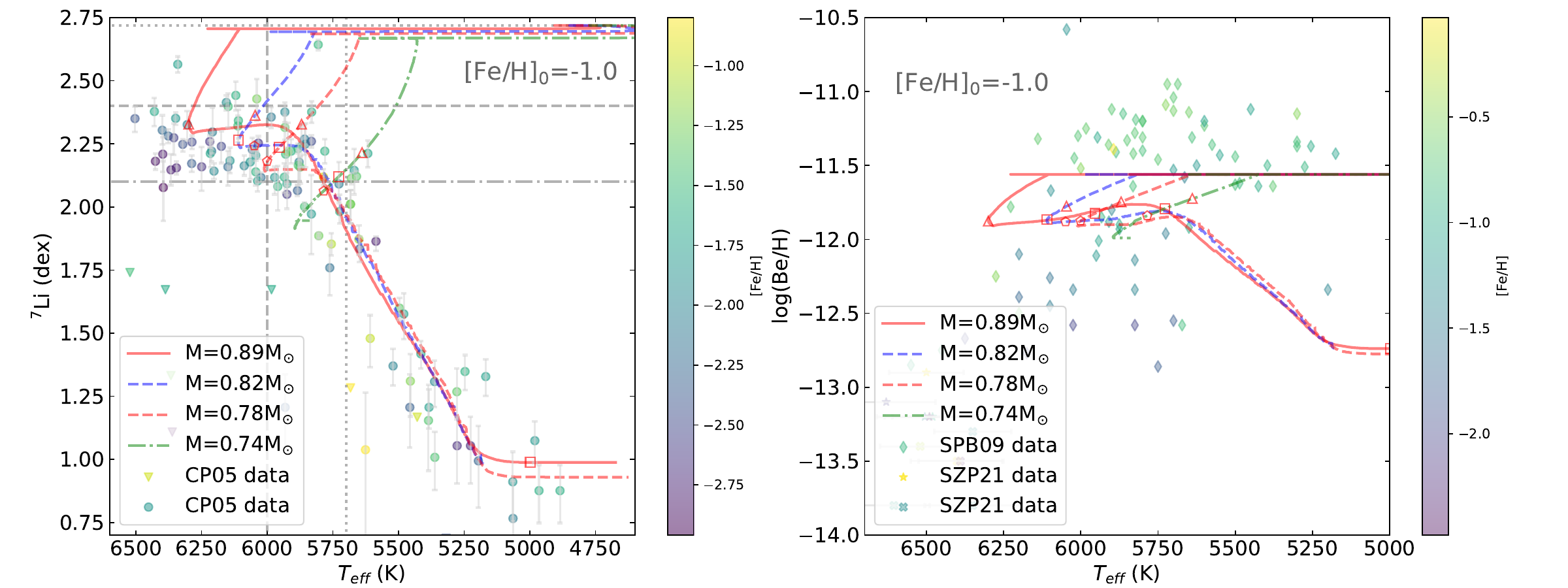}
\includegraphics[angle=0, scale=0.35]{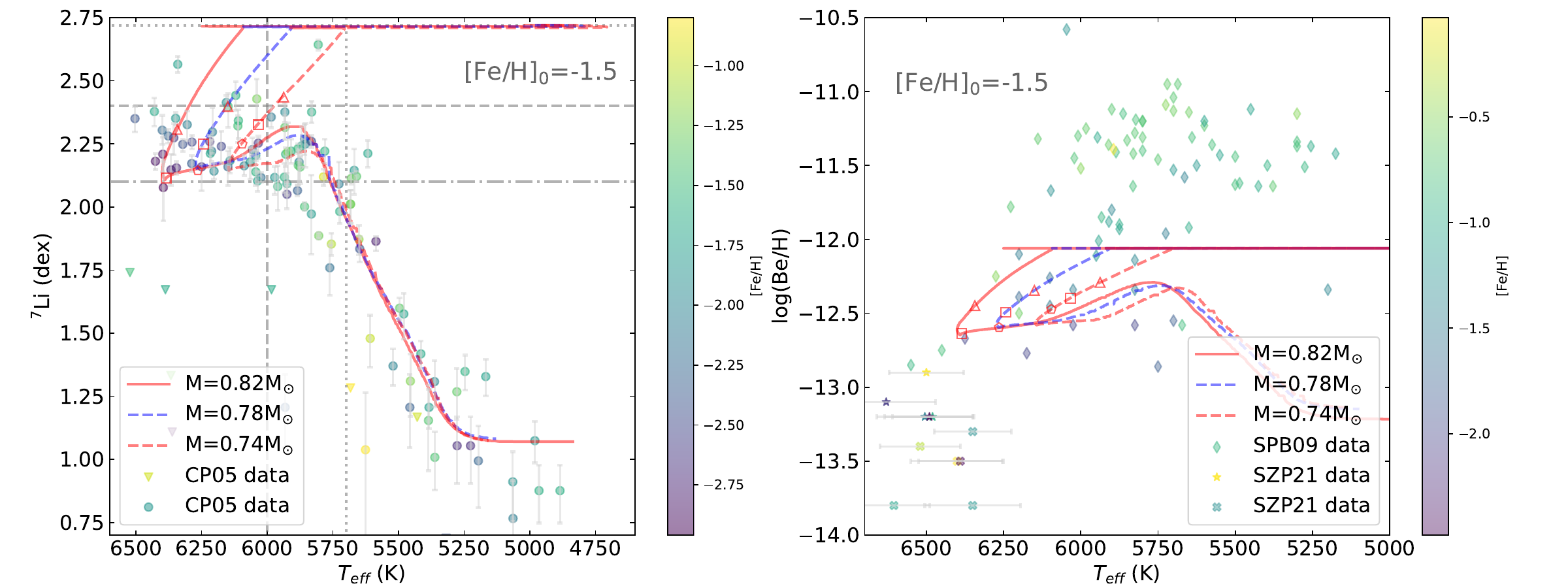}
\includegraphics[angle=0, scale=0.35]{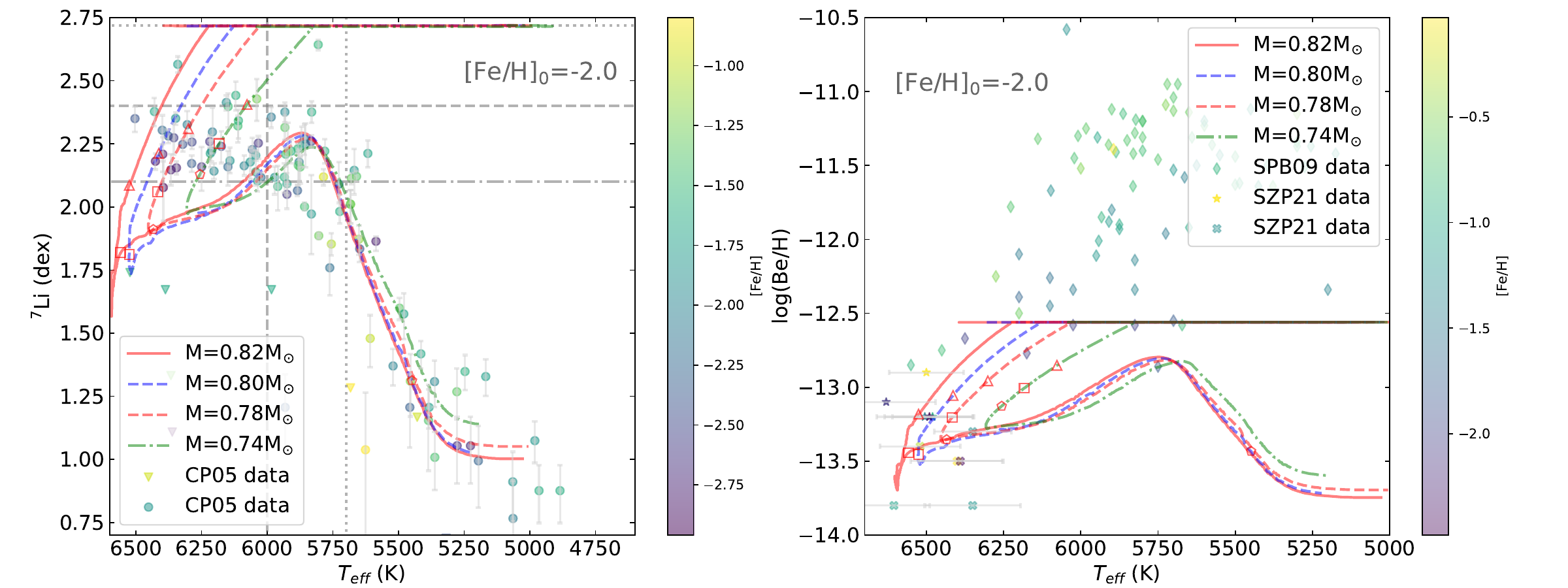}
\caption{Left panels: surface lithium abundance as a function of \teff{} for different models.
The meanings of symbols are same as those in Figure \ref{fig1}.
Right panels: surface beryllium abundance as a function of \teff{} for different models. These models
have a higher initial velocity, as well as C, O, Mg etc enhancement. Filled x represents the observed
Be upper limits determined by \citet{smil21}, while diamond and star denote observational data given
by \citet{smil09} and \citet{smil21}, respectively.
\label{fig11}}
\end{figure*}

\begin{figure*}[ht!]
\plotone{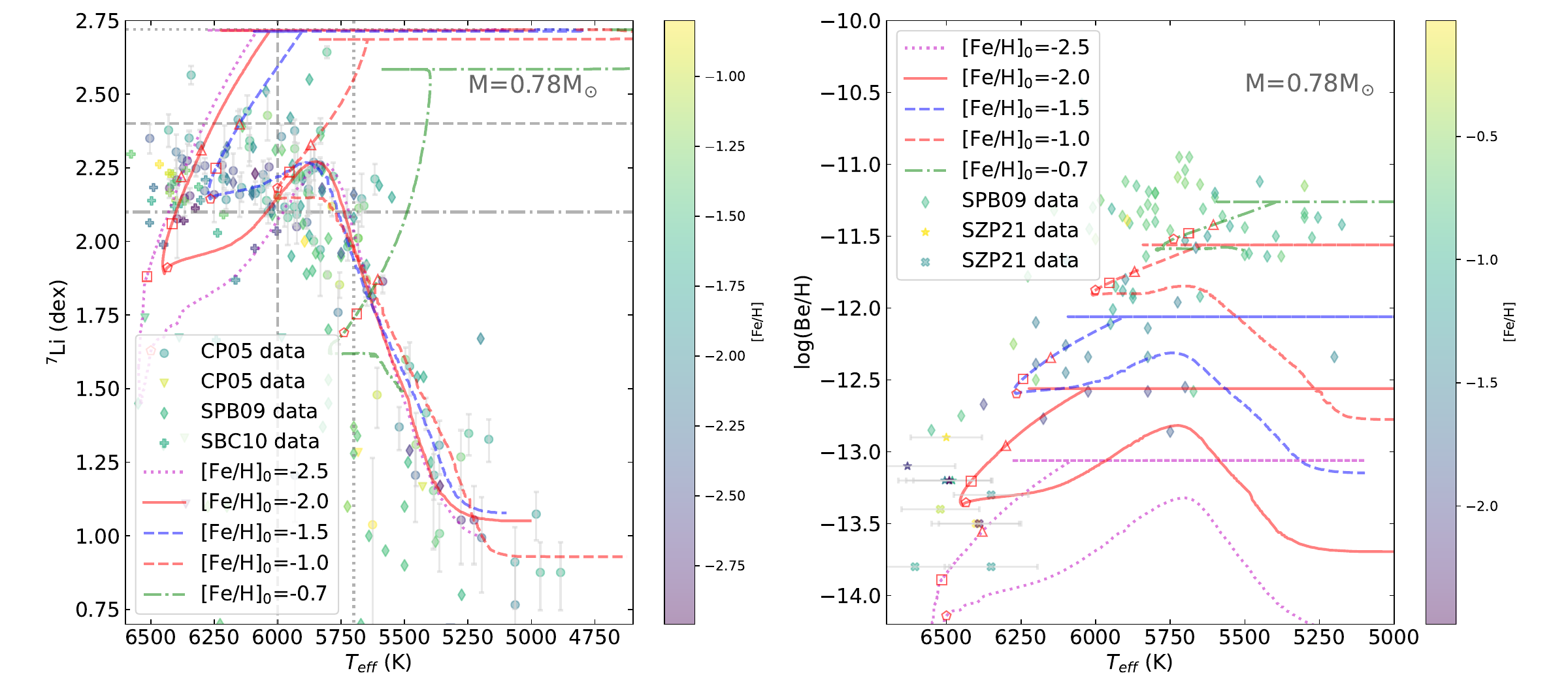}
\plotone{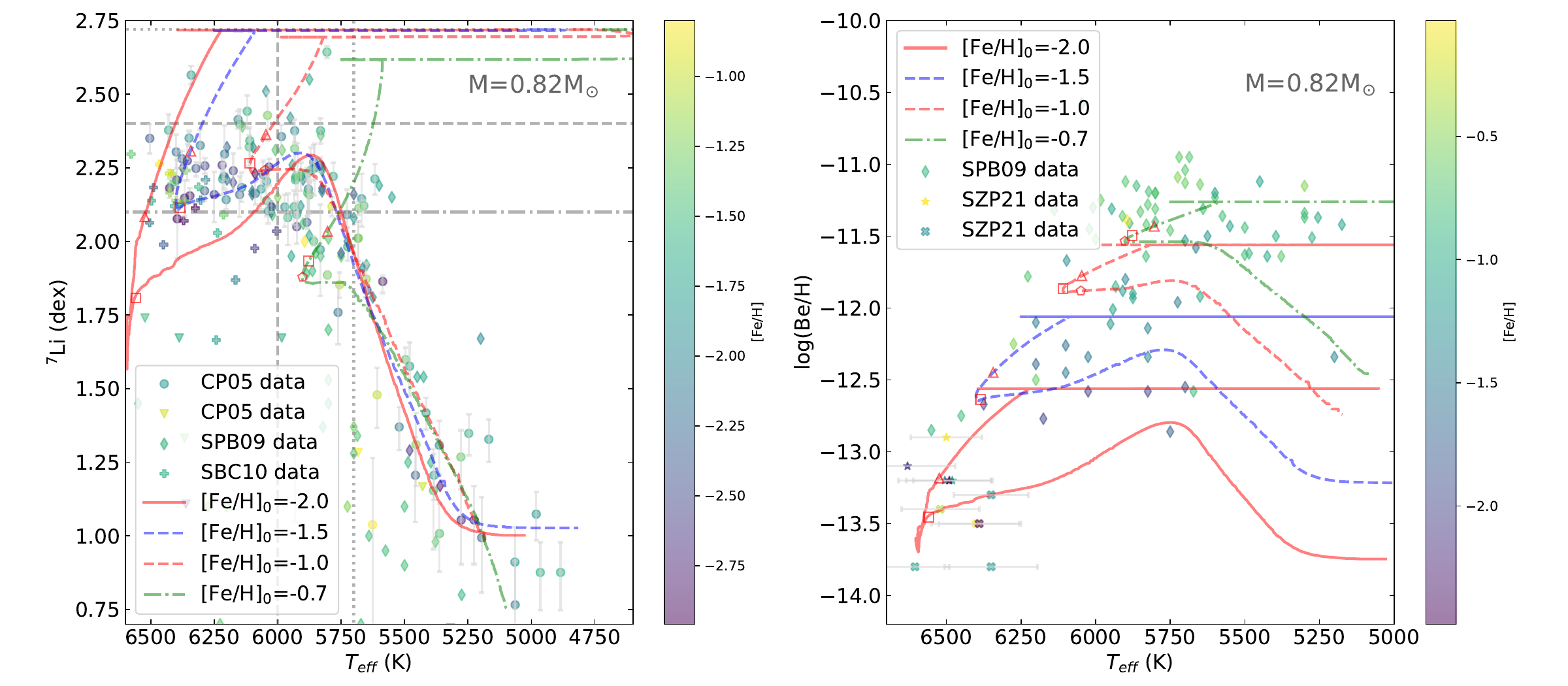}
\caption{Same as Figure \ref{fig11} but for models with the same mass but different metallicities.
Bold solid plus represents the Li abundances determined by \citet{sbor10}. The parameters of the models
with age = 8, 11, and 13 Gyr are listed in Table \ref{tab1}. Those of corresponding nonrotating models are
given in Table \ref{tab2}.
\label{fig12}}
\end{figure*}

These enhancements incorporated into the models shown in Figures \ref{fig11} and \ref{fig12},
including their effects on the opacity tables. Figure \ref{fig11} shows that the distributions of Li
and Be abundances in the $\alpha$-enhanced mixture models with ages between 8 and 13 Gyr are in slightly
better agreement with the observations than those in the solar-scaled mixture models (Figures \ref{fig1},
\ref{fig4}, \ref{fig7}). For example, at [Fe/H]$_{0} = -1.5$, the $\alpha$-enhanced mixture models
predict \ali{} values approximately 2.1–2.4 dex, whereas the solar-scaled mixture models predict \ali{}
mainly in the range of about 2.0--2.4 dex (see Figure \ref{fig4}). Furthermore, the Li abundances of
dwarfs with \teff{} $< 6000$ K are more readily reproduced by the $\alpha$-enhanced mixture models
(see the top panels of Figure \ref{fig11}). These improvements arise from $\alpha$-enhancement rather than
from an increase in the rotation rate. However, these enhancements do not change our main conclusions.

Primordial Li originates from BBN, whereas Be is a pure product of cosmic ray spallation.
Therefore, the initial Be abundance is expected to be a function of time or [Fe/H]; however,
this relation remains uncertain. \citet{smil09, smil21} measured Be abundances for a large
sample of halo and thick-disk stars and derived the following Be-Fe relation for halo stars:
\begin{equation}
 \log(\mathrm{Be/H}) = (-10.76 \pm 0.15) + (0.97 \pm 0.10) [\mathrm{Fe/H}].
\end{equation}
We assume that the initial Be and B abundances can be parameterized as
\begin{equation}
 \log(\mathrm{Be/H}) = \log(\mathrm{Be/H})_{0} + k_{0} [\mathrm{Fe/H}]
 \label{beeq}
\end{equation}
and
\begin{equation}
 \log(\mathrm{B/H}) = \log(\mathrm{B/H})_{0} + k_{0} [\mathrm{Fe/H}],
 \label{beq}
\end{equation}
respectively, where $k_{0} = 1$ is adopted based on the fit of \citet{smil09}. The values of
$\log(\mathrm{Be/H})_{0}$ and $\log(\mathrm{B/H})_{0}$ are set to $-10.56$ and $-9.15$, respectively,
as determined from the solar abundances of \citet{aspl21}.

The right panels of Figures \ref{fig11} and \ref{fig12} compare the predicted and observed $\log(\mathrm{Be/H})$
values, showing that the predicted Be-abundance distributions agree well with the observations.
For a given [Fe/H], the theoretical models predict the presence of a Be plateau with a mild negative
slope in $\log(\mathrm{Be/H})$ versus \teff{}. However, the predicted plateau decreases in abundance
and shifts toward higher effective temperatures with decreasing metallicity, while becoming increasingly
dispersed. Consequently, halo stars do not exhibit a clearly defined Be plateau, although such a feature
may be detectable in star clusters.

\citet{sbor10} found that the Spite plateau undergoes a meltdown at metallicities below $\sim -3$,
while \citet{smil21} showed that the Be-abundance distribution of stars with [Fe/H] $\sim -3$ deviates
from that of stars with higher metallicities (see Figure \ref{fig12} or the figure 2 of \citealt{smil21}).
The stars studied by \citet{sbor10} and \citet{smil09, smil21} are shown in Figure \ref{fig12}.
The left panels of Figure \ref{fig12} show that the Li abundances reported by \citet{sbor10} are well
reproduced by the $\alpha$-enhanced mixture rotating models, and that the Be abundances measured
by \citet{smil09, smil21} are also successfully reproduced.

Table \ref{tab1} lists the Li, Be, along with other parameters, of the rotating models at ages of 8,
11, and 13 Gyr. The corresponding nonrotating models are presented in Table \ref{tab2}. The rotating
models that exhibit both the Spite plateau meltdown and deviations in Be abundances have [Fe/H]
$\lesssim -2.5$ and effective temperatures mainly in the range 6500 K $\gtrsim$ \teff{} $\gtrsim$ 6350 K
(SGB stars exhibiting the Spite plateau meltdown have a lower \teff{}), consistent with the characteristics
of the samples studied by \citet{sbor10} and \citet{smil09, smil21}.

Both the Spite plateau meltdown and the deviation in Be abundances can be reproduced by the same evolutionary
models, indicating that they likely originate from a common mechanism--namely, the presence of a shallower
CZ in these metal-poor stars. Consequently, rotational mixing is insufficient to counteract gravitational
settling. In contrast, stars with higher metallicities generally possess deeper CZs; as a result, gravitational
settling is weaker and more effectively counteracted by rotational mixing. Therefore, the Li and Be abundance
distributions of these metal-poor stars deviate from those of stars with higher metallicities and lower effective
temperatures (see Figure \ref{fig12}).

\subsection{Comparison with the Results of \citealt{nguy25a}}

Based on the grid of stellar models constructed by \citet{nguy25b}, in which the envelope overshoot efficiency
depends on the initial mass and evolutionary stage---i.e. different values of the envelope overshoot efficiency
parameter are adopted for the pre-MS and post-MS phases---\citet{nguy25a} demonstrated that both the Spite plateau
and the second \ali{} plateau observed in early RGB stars can be reproduced by their models. Both the models
of \citet{nguy25a} and our models are able to reproduce the Spite plateau and the second \ali{} plateau.
Moreover, our models also reproduce the Be distributions of the sample reported by \citet{smil09, smil21}.
However, convective overshoot is not included in our models; instead, we consider the effects of rotation
and magnetic fields. The parameters of our rotating models are calibrated to the Sun.

Both \citet{nguy25b} and our work include diffusion and show that MS stars reach a minimum Li abundance
when diffusion operates at maximum efficiency (see Figure 3 of \citealt{nguy25a}), i.e. when the CZ depth
reaches a minimum. This suggests that the effects of diffusion on elements in their models are similar to
those in ours.

The second \ali{} plateau mainly arises from convective dilution during the late SGB stage and the rapid
change in CZ depth during the early RGB stage, rather than from rotational effects. Therefore, both the models of
\citet{nguy25a} and our models are able to predict the existence of a second \ali{} plateau.

In the models of \citet{nguy25a}, more massive stars exhibit higher Li abundances than lower-mass stars
(see their Figure 3). Because the effective temperature increases with stellar mass, the Spite plateau
predicted by their models should therefore rise with effective temperature. This behavior is
inconsistent with both our results and the observations of \citet{lind09}, who found that hot dwarfs
are indeed slightly more Li-poor than cooler dwarfs, unless a larger overshoot---sufficient to
transport Li into regions of efficient nuclear burning---is assumed for their more massive stars.

Moreover, \citet{nguy25a} included Li production from nova explosions in binary systems, which allows
their models to reproduce the observed increase in Li abundance with metallicity at [Fe/H] $> -0.7$.
This contribution is not included in our models; therefore, the corresponding increase in Li abundance
with metallicity cannot be explained within our framework. The Li distributions at [Fe/H] $> -0.7$ may
thus warrant further detailed investigation in future work.

\section{Discussion and Summary} \label{sec4}

\subsection{Discussion}

In our calculations, the initial metal abundance is treated as a free parameter, while the initial
hydrogen and helium abundances are determined by Equations (\ref{xi}) and (\ref{yi}), respectively.
However, they must satisfy the condition $X_{0}+Y_{0}+Z_{0}=1$. The mixing efficiencies
($f_{c}$ and $f_{cm}$) are calibrated to the Sun \citep{yang25}, and the initial Li abundance is
set by BBN. However, for halo stars, the initial abundances of Be and B are unknown; therefore,
we adopt the solar initial abundances. As a result, the Be and B profiles shown in Figures \ref{fig8}
and \ref{fig9} may be shifted, but this does not affect our conclusions, such as the dilution of Be.

Moreover, by assuming that the initial Be and B abundances depend on [Fe/H], we can simultaneously
reproduce the observed Li and Be distributions of halo stars, including the samples of \citet{sbor10} and
\citet{smil21}. This suggests that the differences between the Li and Be distributions in halo stars
arise from both their distinct nucleosynthetic origins and stellar evolutionary effects. The dependence
of Be on [Fe/H] further indicates that large spectroscopic surveys of Be will be valuable for advancing
our understanding of Galactic chemical evolution.

In the solar-scaled mixture models with [Fe/H]$_{0} = - 2.0$ dex and $M \geq 0.78$ \dsm{},
the parameter $f_{cm}$ was increased from its default value of 0.0002 to 0.0005. The parameter
$f_{cm}$ characterizes the efficiency of magnetic-instability-induced mixing. In the Sun, rotational
mixing enhances Li depletion; therefore, increasing $f_{cm}$ leads to stronger Li depletion \citep{yang25}.
In contrast, for stars exhibiting the Spite plateau, rotational mixing mitigates Li depletion. Consequently,
the Li abundances of models with $f_{cm}$ = 0.0005 are enhanced by about 0.1 dex during the MS stage.
However, these abundances still remain below the Spite plateau at the MSTO, indicating that gravitational
settling in these stars cannot be fully counteracted by rotational mixing.

Low-mass stars ($M \lesssim 0.76$ \dsm{}) with [Fe/H]$_{0} = - 2.0$ dex have not yet reached
the MSTO, whereas more massive stars ($M = 0.80$ \dsm{}) have already evolved to the RGB by 13 Gyr.
Therefore, only stars with masses around $0.78$ \dsm{}---or \teff{} around 6400 K---exhibit Li
abundances below the Spite plateau.

All models with $\alpha$-enhanced mixtures are computed using the default values of $f_{c}$ and
$f_{cm}$, together with the low-temperature opacity tables of \citet{mari09} and \citet{mari22},
reconstructed for the $\alpha$-enhanced mixture. $\alpha$-element enhancement leads to a higher
metallicity $Z$. Consequently, $\alpha$-enhanced mixture models possess deeper CZs than solar-scaled
mixture models, which can affect the efficiency of gravitational settling. As a result, the predicted
Li and Be abundances are influenced by $\alpha$-enhancement; however, our main conclusions remain unchanged.

In YREC, the settling velocity $w_{s}$ of species $s$ is computed following Equations (39) and
(41) of \citet{thou94}, where the coefficients $A_{p}(s)$, $A_{T}(s)$, and $A_{t}(s)$ are obtained by
solving Equation (21) of \citet{thou94} at each evolutionary time step; that is, $w_{s}$ is determined
by directly solving the Burgers equations and their associated constraints. Although diffusion velocities
can, in principle, be calculated for all elements using the Equation (21), differential settling would
lead to variations in the heavy-element mixture with stellar age and position.

Because the OPAL and low-temperature opacity tables are constructed for a fixed chemical mixture,
any such variation would require reconstruction of the opacity tables. To avoid this complication,
YREC assumes that, during the hydrogen-burning phase, the settling velocities of all heavy elements
are proportional to that of Fe, such that the total metallicity $Z$ varies with stellar age and
position while the elemental mixture remains unchanged. Under this assumption, the resulting
gravitational settling timescale of heavy elements in the CZ, including Li, scales with the
CZ mass, in agreement with \citet{mich86}.

Moreover, we did not calculate the evolutionary models with [Fe/H]$_{0}\leqslant 10^{-3}$, since the OPAL
or OP \citep{badn05} opacity tables do not provide coverage for metallicities in the range $10^{-4} < Z <0$,
which prevents a reliable study of stars with such low metallicities.

Magnetic fields play a dominant role in angular momentum transport and material mixing in rotating models
\citep{egge22, yang25}. The CZ mass of the $M = 0.78$ \dsm{} model with [Fe/H]$_{0} = - 2.0$ dex at the MSTO is about
$8\times10^{-5}$ \dsm{}, much lower than the Sun’s CZ mass of roughly $2.5\times10^{-2}$ \dsm{}. Such a small
CZ mass may be insufficient to sustain dynamo processes and magnetic braking. In other words, there may exist
a critical CZ mass (m$_{\mathrm{crit}}$) required for the operation of magnetic fields: when m$_{\mathrm{cz}}<$
m$_{\mathrm{crit}}$, magnetic fields become ineffective, and rotational mixing can be neglected. Consequently,
stars with m$_{\mathrm{cz}} <$ m$_{\mathrm{crit}}$ exhibit very low Li abundances during their MS and early
SGB stages due to strong gravitational settling. Because a higher stellar mass or a lower initial metallicity
leads to a shallower CZ, relatively massive or very metal-poor stars are more likely to exhibit this behavior.
This implies that stars with \teff{} $\gtrsim$ 6400 K could have a different Li distribution from those with
\teff{} $<$ 6400 K.

Li dip was observed between effective temperatures of approximately 6400 K and 6800 K \citep{boes86}.
A warm plateau has been identified in stars with $-1.0 \lesssim$ [Fe/H] $\lesssim -0.5$ and
\teff{} $>$ $\sim$6700 K \citep{gao20}. \citet{gao20} argued that such stars may preserve primordial
Li produced in the early Universe and found that the Spite plateau breaks down at [Fe/H] $\approx -0.5$.
For [Fe/H] $\gtrsim -0.5$, the observed \ali{} increases with [Fe/H] due to Galactic chemical enrichment
\citep{gao20, roma21}. \citet{gao20} further suggested that warm plateau stars, Li dip stars, and Spite
plateau stars are governed by different lithium depletion mechanisms, which deserves further detailed
study in future work.

We apply the same mechanisms responsible for solar Li depletion to explain the cosmological
Li problem. To reproduce the seismically inferred CZ depth of the Sun, a convective overshoot of
$\delta_{\mathrm{ov}}=0.09$ is required \citep{yang25}. However, this convective overshoot is not
included in our models, as it does not affect the results for the Spite plateau. This is because
stars on the Spite plateau have shallow CZs. Their \tbc{} is markedly lower than the effective Li-burning
temperature before the middle of the SGB. Such a small convective overshoot is insufficient to
transport Li into regions where it can be effectively burned. For example, even a larger overshoot
of $\delta_{\mathrm{ov}}=0.2$ does not significantly influence the predicted Spite plateau (see the $M = 0.95$
\dsm{} model in Figure \ref{fig7}). Moreover, such an overshoot does not noticeably alter the evolutionary
tracks in the Hertzsprung–Russell diagram.

However, a convective overshoot of $\delta_{\mathrm{ov}}=0.2$ can markedly affect the Li abundances
of late SGB stars and dwarfs with relatively high metallicities and effective temperatures below 6000 K,
as the \tbc{} in these stars approaches the effective Li-burning temperature. For such stars, even a small
overshoot can transport Li into the layers where it can be efficiently destroyed. Since lower-mass
stars have higher \tbc{}, dwarfs cooler than 6000 K with overshoot exhibit a more pronounced decrease
in Li abundance with decreasing effective temperature compared to those without overshoot (see panel (b)
of Figure  \ref{fig7}). In other words, dwarfs with overshoot display Li-depletion behavior similar to
that of SGB stars.

Furthermore, theoretical calculations do not support a large convective overshoot in the surface CZ
(e.g., $\delta_{\mathrm{ov}}\geq 0.7$; \citealt{fu15}), as such an overshoot must lead to excessive
Li depletion during the SGB and RGB stages, resulting in inconsistencies with the Li distributions
observed in SGB and RGB stars \citep{charb05}, unless an overshoot parameter that varies with
stellar mass and age is adopted, as in \citet{nguy25a}. Therefore, the Li abundances of cool dwarfs
and SGB stars provide a more powerful diagnostic of convective overshoot than those of hotter dwarfs.
Unlike previous studies, convective overshoot is not required in our models to explain the Spite plateau.

Moreover, \citet{bara17} proposed that the penetration depth of overshooting depends on the stellar
rotation rate, with rapid rotation strongly limiting the vertical penetration of the convective plumes.
Furthermore, \citet{cons21} showed that fast rotation can suppress convection. Rotationally dependent
overshooting has been used to explain the observed correlation between rotation and Li depletion in
pre-MS stars \citep{bara17, cons21, dumo21}. However, in this study, we do not consider the suppressive
effect of rapid rotation on convection and rotationally dependent overshooting. Consequently,
we also do not model the evolution of rapid rotators.

If the suppression effect of rapid rotation on convection exists in hot dwarf stars, it would lead to a decrease
in the mass of the CZ. The m$_{\mathrm{cz}}$ could then be lower than m$_{\mathrm{crit}}$. Consequently,
magnetic braking and mixing would not operate in these stars, and their Li-abundance evolution would be
dominated by gravitational settling. As a result, these stars could exhibit faster rotation and lower Li
abundances than those on the Spite plateau---that is, they would appear below the plateau while rotating
rapidly. Detailed observations of such stars can help us better understand these phenomena.

For MS stars with \teff{} $\gtrsim$ 6000 K, \tbc{} is significantly lower than the effective Li-burning
temperature. Rotational mixing partially counteracts gravitational settling. The gravitational settling
timescale of Li in the CZ is proportional to the CZ mass \citep{mich86}. The more massive the star,
the shallower its CZ and the stonger the gravitational settling. As a result, our models predict
that hot dwarfs are slightly more Li-poor than cooler ones. Observations of NGC 6397 confirm this
trend---hot dwarfs are indeed slightly more Li-poor than cooler dwarfs \citep{lind09}---providing clear
evidence for the presence of gravitational settling of Li.

Both \citet{korn07} and \citet{lind09} observed that TO stars in NGC 6397 are slightly more Li-poor than SGB stars
that have not yet undergone dilution. This provides strong evidence that the Li-abundance profile below the CZ of
these stars is nearly flat. If \tbc{} in these stars were very close to the effective Li-burning temperature during
the MS stage, Li could not be preserved below the CZ, and the stars would exhibit a decline in Li abundance with
decreasing effective temperature after the MSTO. Moreover, in the absence of gravitational settling, stars would not show
an increase in Li abundance as they evolve from the MSTO to the middle of the SGB. Conversely, if gravitational settling
occurred without rotational mixing, the stars would display a distinct increase in Li abundance, similar to the NMs
shown in Figure \ref{fig5}. In stars where rotational mixing partially counteracts gravitational settling and \teff{}
$\gtrsim$ 6000 K, the Li-abundance profile below the CZ remains nearly flat. As a result, these stars exhibit a slight
increase in Li abundance from the MSTO to the middle of the SGB. Therefore, the observations of \citet{korn07} and
\citet{lind09} suggest that \tbc{} in these stars is significantly lower than the effective Li-burning temperature,
and that both gravitational settling and rotational mixing operate below their CZs.

Elements heavier than B cannot be burned in dwarf stars. These heavy elements deposited below the CZ can be
preserved during the MS stage and later dredged up as the CZ deepens. Consequently, the abundances of heavy
elements can increase by about 0.2 dex from the MSTO to the RGB stage. During the RGB stage, the CNO-cycle
reactions---$^{12}$C($p, \gamma$)$^{13}$N and $^{13}$C($p, \gamma$)$^{14}$N---cause the N abundance to exceed
its initial value, leading to N enrichment (see Figure \ref{fig10}). This enrichment becomes more pronounced
with increasing initial metallicity and stellar mass. Observational constraints on this enrichment
would help us better understand the chains of the CNO cycle. In the early RGB stage, we neglect the effects
of the CNO cycle on the heavy-element mixture.

Because Li is strongly diluted during the late SGB stage, the observed anti-correlation between Li and heavy
elements in SGB stars could originate from stellar evolution; at the very least, this effect cannot be neglected.
Moreover, since Be and B dilutions occur at lower effective temperature due to their higher burning temperatures,
measurements of Be and B abundances in stars can help clarify whether the observed Li–Na anticorrelation in SGB
stars arises from external pollution or from stellar evolutionary processes.

The sample of studied by \citet{smil09, smil21} includes many stars with [Fe/H] $\geq -0.7$. In contrast,
we computed the evolution only for stars with initial [Fe/H] $\leq -0.7$. As a result, our models do not cover
these relatively high-metallicity stars (see top-right panel of Figure \ref{fig11}, where many stars have
$A(\mathrm{Be})$ values higher than those predicted by our models). Stars with [Fe/H] $\geq -0.7$
are influenced by Galactic chemical enrichment \citep{gao20, roma21}. Therefore, the Be deviation reported by
\citet{smil21} may result from both Galactic chemical enrichment and stellar Be depletion.
This issue deserves more detailed investigation in future work.

\subsection{Summary}

In this work, we employed gravitational settling, diffusion, rotation, and magnetic fields to explain the
Spite plateau. The initial hydrogen and helium abundances are determined by Equations (\ref{xi}) and (\ref{yi}),
respectively, which are deduced from the primordial helium abundance inferred from observations and the solar
$Y_{0}$ and $Z_{0}$. The initial metal abundance is a free parameter, while the initial Li abundance is set by
BBN. We computed stellar evolutionary models with [Fe/H]$_{0}$ ranging from $-0.7$ to $-2.0$, corresponding to
[Fe/H] values of approximately $-0.70$ to $-2.75$ during the MS stage, consistent with the metallicity range
of the sample studied by \citet{charb05}. The mixing-length parameter $\alpha_{\mathrm{MLT}}$ is calibrated to
the Sun and assumed to remain constant.

we also include the effects of a tachocline in the rotating models, following \citet{yang25}. The tachocline width
is assumed to be $0.05R$, consistent with the seismically inferred value for the Sun \citep{charb99}. The parameters
governing rotation and magnetic fields, including angular momentum transport and chemical mixing, are calibrated
to the Sun. We compute the evolution of nonrotating, slowly rotating, and moderately rotating models.
Fast rotation and the suppressive effects of rapid rotation on convection and overshooting are not included in
this study. These mechanisms may play a more significant role in cooler dwarfs and warrant further investigation.

The more massive the star or the lower its initial metallicity, the shallower its CZ becomes. Dwarfs with \teff{}
$\gtrsim$ 5900 K possess shallow CZs. The gravitational-settling timescale of Li in the CZ is proportional to the CZ
mass, and the CZ depth also decreases with increasing age or \teff{} prior to the MSTO. Thus, in NMs, gravitational
settling results in Li abundance rapidly decreasing with increasing \teff{} or age before the MSTO. Because \tbc{}
in these stars is markedly lower than the effective Li-burning temperature, Li can be preserved below the CZ.
Gravitational settling produces a negative Li-abundance gradient beneath the CZ. Therefore, as NMs evolve from
the MSTO to the middle of the SGB, their Li abundances increase rapidly due to dredge-up associated with CZ deepening.
Consequently, the Li abundances predicted by NMs cannot account for the Spite plateau.

In rotating models, rotational mixing---including the effects of magnetic fields---partially counteracts
gravitational settling. Consequently, rotation and magnetic fields mitigate Li depletion and nearly smooth
out the Li gradient below the CZ during the MS stage of these stars. As a result, the Li abundances of MS
stars with ages between about 8 and 13 Gyr remain mostly within 2.0--2.4 dex. From the MSTO to the middle of
the SGB, Li abundances predicted by RMs increase only slightly with decreasing \teff{} as the CZ deepens
due to the nearly flat Li profile. Consequently, Li abundances in RMs with \teff{} between approximately
6400 and 5900 K and ages between about 8 and 13 Gyr generally fall within 2.0--2.4 dex, forming a Li plateau.
The plateau exhibits a slight negative slope in the \ali{}-\teff{} plane. However, the dependence of \ali{}
on \teff{} weakens with increasing metallicity and almost vanishes in stars with [Fe/H]$_{0}=-0.7$.
At around 6400 K, very metal-poor stars or fast rotators are likely to exhibit Li abundances below the Spite
plateau due to their shallow CZs.

Unlike hot dwarfs with \teff{} $\gtrsim$ 5900 K, cooler dwarfs have a deeper CZ, such as the Sun. Their \tbc{}
is close to the effective Li-burning temperature during the MS stage. Rotational mixing transports Li into hotter
layers where it can be efficiently burned. Thus, rotational mixing enhances Li depletion in these stars.
The Li abundances of these stars decrease with decreasing mass and are easily affected by convective overshoot.
Therefore, they can serve as a powerful diagnostic of convective overshoot.

From the middle of the SGB to the RGB, variations in CZ depth dominate the changes in Li, Be, and B abundances.
The rapid deepening of the CZ leads to the dilution of Li, Be, and B. Li dilution occurs at about 5900 K,
whereas Be and B dilution takes place at lower temperatures, reflecting their different burning temperatures.
Moreover, rotating models predict another Li plateau with \ali{} $\approx 1.0$ for RGB stars with \teff{}
$\lesssim$ 5200 K.

The rotating models including gravitational settling, diffusion, rotation, and magnetic fields, predict
that Li abundances of metal-poor stars with ages between approximately 8 and 13 Gyr and \teff{} between
6400 and 5900 K generally fall within 2.0--2.4 dex, followed by a sharp decline in Li abundance down to
\teff{} $\approx$ 5200 K. This trend agrees well with the observations reported by \citet{charb05}. The
predicted \ali{} for RGB stars with \teff{} $\lesssim$ 5200 K is about 1.0 dex, consistent with the measurements
of \citet{mucc22}. The predicted Li plateau exhibits a slight negative slope in \ali{} versus \teff{}.
Furthermore, the Li abundance on the plateau increases modestly with metallicity, while the dependence
of \ali{} on \teff{} weakens as metallicity increases. This can be attributed to the fact that the CZ depth
depends on both stellar mass and metallicity, and that the gravitational settling timescale of Li in the CZ
is proportional to the CZ mass. These results suggest that the Spite plateau arises from the combined effects
of variations in CZ depth, gravitational settling, diffusion, rotational mixing, and magnetic fields.
In hot dwarf stars with \teff{} $>$ 5900 K, rotational and magnetic mixing partially counteract gravitational
settling and nearly smooth out the Li gradient below the CZ during the MS stage, causing the Li abundances
to remain mostly within 2.0--2.4 dex. Finally, the adopted initial Li abundance of 2.72 dex agrees well with
the BBN prediction based on WMAP and Planck data, implying that the cosmological Li problem originates
from stellar Li depletion. The rotational and magnetic mixing are calibrated to reproduce the Sun’s Li depletion,
suggesting that the Li depletions of the Sun and halo stars are governed by the same physical mechanisms.
The Sun’s lower Li abundance compared with the Spite plateau can be attributed to its different internal structure,
namely its deeper CZ. Moreover, The distributions of Li reported by \citet{sbor10} and Be measured by
\citet{smil09, smil21} of halo stars are also reproduced by rotating models.

\begin{acknowledgments}
Authors thank the anonymous referee for helpful comments that helped the authors signiﬁcantly improve this work,
and acknowledge the support from the NSFC 12573032, 12288102, 12333008, and 11973080. X.M. acknowledges support from
the NSFC and National Key R\&D Program of China (No. 2021YFA1600403), the Strategic Priority Research Program of the Chinese
Academy of Sciences (grant Nos. XDB1160303, XDB1160000), International Centre of Supernovae, Yunnan Key Laboratory
(No. 202302AN360001), Yunnan Fundamental Research Projects (NOs. 202401BC070007 and 202201BC070003),
the Yunnan Revitalization Talent Support Program-Science \& Technology Champion Project (NO. 202305AB350003), and the science
research grants from the China Manned Space Project with grant no. CMS-CSST-2025-A12, CMS-CSST-2025-A13, and CMS-CSST-2025-A15.
All models and data used in this work are available at yangwuming@bnu.edu.cn.
\end{acknowledgments}





\bibliography{sample7}{}

\begin{thebibliography}{}
\expandafter\ifx\csname natexlab\endcsname\relax\def\natexlab#1{#1}\fi
\providecommand{\url}[1]{\href{#1}{#1}}
\providecommand{\dodoi}[1]{doi:~\href{http://doi.org/#1}{\nolinkurl{#1}}}
\providecommand{\doeprint}[1]{\href{http://ascl.net/#1}{\nolinkurl{http://ascl.net/#1}}}
\providecommand{\doarXiv}[1]{\href{https://arxiv.org/abs/#1}{\nolinkurl{https://arxiv.org/abs/#1}}}

\bibitem[{ {Astropy Collaboration} {et~al.}(2013){Astropy Collaboration},
  {Robitaille}, {Tollerud}, {Greenfield}, {Droettboom}, {Bray}, {Aldcroft},
  {Davis}, {Ginsburg}, {Price-Whelan}, {Kerzendorf}, {Conley}, {Crighton},
  {Barbary}, {Muna}, {Ferguson}, {Grollier}, {Parikh}, {Nair}, {Unther},
  {Deil}, {Woillez}, {Conseil}, {Kramer}, {Turner}, {Singer}, {Fox}, {Weaver},
  {Zabalza}, {Edwards}, {Azalee Bostroem}, {Burke}, {Casey}, {Crawford},
  {Dencheva}, {Ely}, {Jenness}, {Labrie}, {Lim}, {Pierfederici}, {Pontzen},
  {Ptak}, {Refsdal}, {Servillat}, \& {Streicher}}]{2013A&A...558A..33A}
{Astropy Collaboration}, {Robitaille}, T.~P., {Tollerud}, E.~J., {et~al.} 2013,
  \bibinfo{title}{{Astropy: A community Python package for astronomy},} \aap,
  558, A33, \dodoi{10.1051/0004-6361/201322068}

\bibitem[{ {Astropy Collaboration} {et~al.}(2018){Astropy Collaboration},
  {Price-Whelan}, {Sip{\H{o}}cz}, {G{\"u}nther}, {Lim}, {Crawford}, {Conseil},
  {Shupe}, {Craig}, {Dencheva}, {Ginsburg}, {VanderPlas}, {Bradley},
  {P{\'e}rez-Su{\'a}rez}, {de Val-Borro}, {Aldcroft}, {Cruz}, {Robitaille},
  {Tollerud}, {Ardelean}, {Babej}, {Bach}, {Bachetti}, {Bakanov}, {Bamford},
  {Barentsen}, {Barmby}, {Baumbach}, {Berry}, {Biscani}, {Boquien}, {Bostroem},
  {Bouma}, {Brammer}, {Bray}, {Breytenbach}, {Buddelmeijer}, {Burke},
  {Calderone}, {Cano Rodr{\'\i}guez}, {Cara}, {Cardoso}, {Cheedella}, {Copin},
  {Corrales}, {Crichton}, {D'Avella}, {Deil}, {Depagne}, {Dietrich}, {Donath},
  {Droettboom}, {Earl}, {Erben}, {Fabbro}, {Ferreira}, {Finethy}, {Fox},
  {Garrison}, {Gibbons}, {Goldstein}, {Gommers}, {Greco}, {Greenfield},
  {Groener}, {Grollier}, {Hagen}, {Hirst}, {Homeier}, {Horton}, {Hosseinzadeh},
  {Hu}, {Hunkeler}, {Ivezi{\'c}}, {Jain}, {Jenness}, {Kanarek}, {Kendrew},
  {Kern}, {Kerzendorf}, {Khvalko}, {King}, {Kirkby}, {Kulkarni}, {Kumar},
  {Lee}, {Lenz}, {Littlefair}, {Ma}, {Macleod}, {Mastropietro}, {McCully},
  {Montagnac}, {Morris}, {Mueller}, {Mumford}, {Muna}, {Murphy}, {Nelson},
  {Nguyen}, {Ninan}, {N{\"o}the}, {Ogaz}, {Oh}, {Parejko}, {Parley}, {Pascual},
  {Patil}, {Patil}, {Plunkett}, {Prochaska}, {Rastogi}, {Reddy Janga},
  {Sabater}, {Sakurikar}, {Seifert}, {Sherbert}, {Sherwood-Taylor}, {Shih},
  {Sick}, {Silbiger}, {Singanamalla}, {Singer}, {Sladen}, {Sooley},
  {Sornarajah}, {Streicher}, {Teuben}, {Thomas}, {Tremblay}, {Turner},
  {Terr{\'o}n}, {van Kerkwijk}, {de la Vega}, {Watkins}, {Weaver}, {Whitmore},
  {Woillez}, {Zabalza}, \& {Astropy Contributors}}]{2018AJ....156..123A}
{Astropy Collaboration}, {Price-Whelan}, A.~M., {Sip{\H{o}}cz}, B.~M., {et~al.}
  2018, \bibinfo{title}{{The Astropy Project: Building an Open-science Project
  and Status of the v2.0 Core Package},} \aj, 156, 123,
  \dodoi{10.3847/1538-3881/aabc4f}

\bibitem[{ {Astropy Collaboration} {et~al.}(2022){Astropy Collaboration},
  {Price-Whelan}, {Lim}, {Earl}, {Starkman}, {Bradley}, {Shupe}, {Patil},
  {Corrales}, {Brasseur}, {N{\"o}the}, {Donath}, {Tollerud}, {Morris},
  {Ginsburg}, {Vaher}, {Weaver}, {Tocknell}, {Jamieson}, {van Kerkwijk},
  {Robitaille}, {Merry}, {Bachetti}, {G{\"u}nther}, {Aldcroft},
  {Alvarado-Montes}, {Archibald}, {B{\'o}di}, {Bapat}, {Barentsen},
  {Baz{\'a}n}, {Biswas}, {Boquien}, {Burke}, {Cara}, {Cara}, {Conroy},
  {Conseil}, {Craig}, {Cross}, {Cruz}, {D'Eugenio}, {Dencheva}, {Devillepoix},
  {Dietrich}, {Eigenbrot}, {Erben}, {Ferreira}, {Foreman-Mackey}, {Fox},
  {Freij}, {Garg}, {Geda}, {Glattly}, {Gondhalekar}, {Gordon}, {Grant},
  {Greenfield}, {Groener}, {Guest}, {Gurovich}, {Handberg}, {Hart},
  {Hatfield-Dodds}, {Homeier}, {Hosseinzadeh}, {Jenness}, {Jones}, {Joseph},
  {Kalmbach}, {Karamehmetoglu}, {Ka{\l}uszy{\'n}ski}, {Kelley}, {Kern},
  {Kerzendorf}, {Koch}, {Kulumani}, {Lee}, {Ly}, {Ma}, {MacBride}, {Maljaars},
  {Muna}, {Murphy}, {Norman}, {O'Steen}, {Oman}, {Pacifici}, {Pascual},
  {Pascual-Granado}, {Patil}, {Perren}, {Pickering}, {Rastogi}, {Roulston},
  {Ryan}, {Rykoff}, {Sabater}, {Sakurikar}, {Salgado}, {Sanghi}, {Saunders},
  {Savchenko}, {Schwardt}, {Seifert-Eckert}, {Shih}, {Jain}, {Shukla}, {Sick},
  {Simpson}, {Singanamalla}, {Singer}, {Singhal}, {Sinha}, {Sip{\H{o}}cz},
  {Spitler}, {Stansby}, {Streicher}, {{\v{S}}umak}, {Swinbank}, {Taranu},
  {Tewary}, {Tremblay}, {de Val-Borro}, {Van Kooten}, {Vasovi{\'c}}, {Verma},
  {de Miranda Cardoso}, {Williams}, {Wilson}, {Winkel}, {Wood-Vasey}, {Xue},
  {Yoachim}, {Zhang}, {Zonca}, \& {Astropy Project
  Contributors}}]{2022ApJ...935..167A}
{Astropy Collaboration}, {Price-Whelan}, A.~M., {Lim}, P.~L., {et~al.} 2022,
  \bibinfo{title}{{The Astropy Project: Sustaining and Growing a
  Community-oriented Open-source Project and the Latest Major Release (v5.0) of
  the Core Package},} \apj, 935, 167, \dodoi{10.3847/1538-4357/ac7c74}

\bibitem[{E. {Bertin} \& S. {Arnouts}(1996){Bertin} \&
  {Arnouts}}]{1996A&AS..117..393B}
{Bertin}, E., \& {Arnouts}, S. 1996, \bibinfo{title}{{SExtractor: Software for
  source extraction.},} \aaps, 117, 393, \dodoi{10.1051/aas:1996164}

\bibitem[{L. {Corrales}(2015){Corrales}}]{2015ApJ...805...23C}
{Corrales}, L. 2015, \bibinfo{title}{{X-Ray Scattering Echoes and Ghost Halos
  from the Intergalactic Medium: Relation to the Nature of AGN Variability},}
  \apj, 805, 23, \dodoi{10.1088/0004-637X/805/1/23}

\bibitem[{G.~J. {Ferland} {et~al.}(2013){Ferland}, {Porter}, {van Hoof},
  {Williams}, {Abel}, {Lykins}, {Shaw}, {Henney}, \&
  {Stancil}}]{2013RMxAA..49..137F}
{Ferland}, G.~J., {Porter}, R.~L., {van Hoof}, P.~A.~M., {et~al.} 2013,
  \bibinfo{title}{{The 2013 Release of Cloudy},} \rmxaa, 49, 137.
\newblock \doarXiv{1302.4485}

\bibitem[{R.~J. {Hanisch} \& C.~D. {Biemesderfer}(1989){Hanisch} \&
  {Biemesderfer}}]{1989BAAS...21..780H}
{Hanisch}, R.~J., \& {Biemesderfer}, C.~D. 1989, in \baas, 780

\bibitem[{L. {Lamport}(1994){Lamport}}]{lamport94}
{Lamport}, L. 1994, {LaTeX: A Document Preparation System}, 2nd edn.
  (Addison-Wesley Professional)

\end{thebibliography}
\bibliographystyle{aasjournal}



\begin{longrotatetable}
\begin{deluxetable*} {lcccccccccccc}
\bf{
\tablecaption{Parameters of Rotating Models.
\label{tab1}}
\tablehead{
\colhead{Mass} & \colhead{Age} & \colhead{[Fe/H]} & \teff{} &\colhead{Velocity} & $T_{\mathrm{BCZ}}$ &\colhead{$A$(Li)} & \colhead{$A$(Be) or} & \colhead{$A$(B) or} & \colhead{$A$(C)} &\colhead{$A$(N)} &\colhead{$A$(O)} & \colhead{$\alpha$-enhanced$^{a}$} \\
\colhead{\dsm{}} & \colhead{Gyr} & & K &\colhead{km s$^{-1}$} & 10$^{6}$ K & & log(Be/H) & log(B/H) &  &  &  &
}
\startdata
  0.65 &  0 & -1.00 &  4727 & 5.4 & ...$^{b}$ & 2.72 & 1.44 & 2.85 & 7.55 & 6.97 & 7.76 & No \\
       &  0.06$^{c}$ & -1.00 &  5028 & 3.9 & 2.14 & 2.65 & 1.44 & 2.85 & 7.55 & 6.97 & 7.76 & No \\
       &  8    & -1.08 &  5175 & 0.4 & 2.05 & 2.24 & 1.29 & 2.70 & 7.49 & 6.91 & 7.70 & No \\
       & 11   & -1.11 &  5246 & 0.3 & 2.02 & 2.17 & 1.23 & 2.64 & 7.46 & 6.88 & 7.67 & No \\
       & 13   & -1.13 &  5295 & 0.3 & 1.99 & 2.12 & 1.18 & 2.60 & 7.44 & 6.86 & 7.65 & No \\
       & 27.1$^{d}$ & -1.25 &  5678 & 0.2 & 1.63 & 1.83 & 0.90 & 2.31 & 7.32 & 6.74 & 7.53 & No \\
       & 28.3$^{e}$ & -1.25 &  5658 & 0.2 & 1.65 & 1.83 & 0.90 & 2.31 & 7.32 & 6.74 & 7.53 & No \\
       & 29.5$^{f}$ & -1.22 &  5444 & 0.2 & 1.83 & 1.59 & 1.00 & 2.48 & 7.38 & 6.80 & 7.59 & No \\
       & 30.0$^{g}$ & -1.14 &  5188 & 0.1 & 2.52 & 1.07 & 0.49 & 2.51 & 7.46 & 6.88 & 7.67 & No \\
\hline
 0.74 &  0  & -1.0  & 4786 & 6.7  & ... & 2.72 & 1.44 & 2.85 & 7.55 & 6.97 & 7.76  & No \\
      &  8  & -1.11 & 5747 & 0.63 & 1.69 & 2.40 & 1.25 & 2.66 & 7.46 & 6.88 & 7.67  & No\\
      &  11 & -1.15 & 5840 & 0.53 & 1.51 & 2.31 & 1.16 & 2.57 & 7.43 & 6.85 & 7.64  & No   \\
      &  13 & -1.18 & 5898 & 0.47 & 1.46 & 2.25 & 1.11 & 2.52 & 7.40 & 6.82 & 7.61  & No   \\
\hline
 0.78 &  0  & -1.0  & 4808 & 7.1  & ... & 2.72 & 1.44 & 2.85 & 7.55 & 6.97 & 7.76  & No \\
      &  8  & -1.14 & 5958 & 0.72 & 1.44 & 2.41 & 1.20 & 2.61 & 7.44 & 6.86 & 7.65  & No\\
      &  11 & -1.19 & 6048 & 0.65 & 1.21 & 2.32 & 1.11 & 2.52 & 7.40 & 6.82 & 7.61  & No   \\
      &  13 & -1.23 & 6091 & 0.61 & 1.07 & 2.25 & 1.04 & 2.45 & 7.37 & 6.79 & 7.58  & No   \\
\hline
 0.82 &  0  & -1.0  & 4840 & 7.1  & ... & 2.72 & 1.44 & 2.85 & 7.55 & 6.97 & 7.76 & No \\
      &  0.037$^{c}$ & -1.00 &  5929 & 5.3 & 1.61 & 2.71 & 1.44 & 2.85 & 7.55 & 6.97 & 7.76 & No \\
      &  8  & -1.18 & 6151 & 0.84 & 1.09 & 2.39 & 1.15 & 2.56 & 7.42 & 6.84 & 7.63 & No \\
      &  11$^{d}$   & -1.26 & 6210 & 0.73 & 0.83 & 2.26 & 1.02 & 2.43 & 7.34 & 6.79 & 7.58 & No   \\
      & 12.4$^{e}$  & -1.27 &  6129 & 0.7 & 0.82 & 2.25 & 1.02 & 2.43 & 7.36 & 6.78 & 7.57 & No \\
      & 12.97$^{f}$ & -1.18 &  5836 & 0.4 & 1.21 & 2.29 & 1.15 & 2.57 & 7.42 & 6.84 & 7.63 & No \\
      &  13         & -1.16 & 5782 & 0.40 & 1.31 & 2.26 & 1.18 & 2.59 & 7.43 & 6.85 & 7.64 & No   \\
      & 13.4$^{g}$  & -1.03 &  5254 & 0.2 & 2.49 & 1.19 & 0.45 & 2.41 & 7.52 & 6.94 & 7.73 & No \\
\hline
  0.74 &  0 & -0.70 &  4571 & 38.0 & 1.06 & 2.72 & -11.26 & -9.85 & 8.04 & 7.28 & 8.54 & Yes \\
       &  8 & -0.77 &  5379 & 0.69 & 2.26 & 1.68 & -11.41 & -9.98 & 7.98 & 7.22 & 8.48 & Yes \\
       & 11 & -0.80 &  5463 & 0.51 & 2.17 & 1.58 & -11.47 & -10.05 & 7.95 & 7.19 & 8.45 & Yes \\
       & 13 & -0.82 &  5520 & 0.45 & 2.15 & 1.50 & -11.52 & -10.10 & 7.93 & 7.17 & 8.43 & Yes \\
\hline
  0.74 &  0 & -1.00 &  4709 & 38.0 & 1.12 & 2.72 & -11.56 & -10.15 & 7.74 & 6.98 & 8.24 & Yes \\
       &  8 & -1.09 &  5639 & 0.67 & 1.95 & 2.21 & -11.72 & -10.31 & 7.67 & 6.91 & 8.17 & Yes \\
       & 11 & -1.12 &  5728 & 0.55 & 1.84 & 2.12 & -11.79 & -10.38 & 7.64 & 6.88 & 8.14 & Yes \\
       & 13 & -1.15 &  5785 & 0.45 & 1.77 & 2.06 & -11.84 & -10.43 & 7.62 & 6.86 & 8.12 & Yes \\
\hline
  0.74 &  0 & -1.50 &  4928 & 38.0 & 1.46 & 2.72 & -12.06 & -10.65 & 7.24 & 6.48 & 7.74 & Yes \\
       &  8 & -1.63 &  5937 & 0.63 & 1.45 & 2.44 & -12.29 & -10.88 & 7.14 & 6.38 & 7.64 & Yes \\
       & 11 & -1.69 &  6034 & 0.53 & 1.29 & 2.33 & -12.40 & -10.99 & 7.09 & 6.33 & 7.59 & Yes \\
       & 13 & -1.73 &  6095 & 0.48 & 1.14 & 2.25 & -12.47 & -11.06 & 7.06 & 6.30 & 7.56 & Yes \\
\hline
  0.74 &  0 & -2.00 &  5056 & 38.0 & 1.45 & 2.72 & -12.56 & -11.15 & 6.74 & 5.98 & 7.24 & Yes \\
       &  8 & -2.17 &  6077 & 0.65 & 1.15 & 2.41 & -12.85 & -11.44 & 6.61 & 5.85 & 7.11 & Yes \\
       & 11 & -2.26 &  6183 & 0.53 & 0.96 & 2.25 & -13.01 & -11.60 & 6.55 & 5.79 & 7.05 & Yes \\
       & 13 & -2.33 &  6256 & 0.49 & 0.79 & 2.13 & -13.13 & -11.72 & 6.50 & 5.73 & 6.99 & Yes \\
\hline
 0.78 &  0  & -0.7  & 4613 & 39.0 & ...  & 2.72 & -11.26 & -9.85 & 8.04 & 7.28 & 8.54  &  Yes  \\
      &  8  & -0.78 & 5605 & 0.79 & 2.10 & 1.87 & -11.42 & -10.00 & 7.97 & 7.21 & 8.47  &  Yes  \\
      &  11 & -0.82 & 5688 & 0.67 & 2.04 & 1.75 & -11.48 & -10.06 & 7.94 & 7.18 & 8.44  &  Yes  \\
      &  13 & -0.84 & 5738 & 0.61 & 1.97 & 1.69 & -11.52 & -10.10 & 7.92 & 7.17 & 8.43  &  Yes  \\
\hline
 0.78 &  0  & -1.0  & 4807 & 39.0 & ...  & 2.72 & -11.56 & -10.15 & 7.72 & 6.96 & 8.22  &  Yes  \\
      &  8  & -1.12 & 5869 & 0.76 & 1.74 & 2.33 & -11.75 & -10.33 & 7.64 & 6.88 & 8.14  &  Yes  \\
      &  11 & -1.17 & 5954 & 0.66 & 1.59 & 2.24 & -11.83 & -10.41 & 7.60 & 6.85 & 8.10  &  Yes  \\
      &  13 & -1.19 & 6000 & 0.62 & 1.46 & 2.18 & -11.88 & -10.46 & 7.58 & 6.82 & 8.08  &  Yes  \\
\hline
 0.78 &  0  & -1.5  & 4909 & 39.0 & ... & 2.72 & -12.06 & -10.65 & 7.24 & 6.48 & 7.74  &  Yes  \\
      &  8  & -1.67 & 6150 & 0.77 & 1.15 & 2.40 & -12.35 & -10.93 & 7.11 & 6.35 & 7.61  &  Yes  \\
      &  11 & -1.76 & 6243 & 0.65 & 0.89 & 2.25 & -12.49 & -11.08 & 7.05 & 6.29 & 7.55  &  Yes  \\
      &  13 & -1.83 & 6264 & 0.61 & 0.72 & 2.14 & -12.60 & -11.19 & 7.01 & 6.25 & 7.51  &  Yes  \\
\hline
 0.78 &  0  & -2.0  & 5089 & 39.0 & ... & 2.72 & -12.56 & -11.15 & 6.74 & 5.98 & 7.24  &  Yes  \\
      &  8  & -2.26 & 6301 & 0.77 & 0.83 & 2.31 & -12.96 & -11.55 & 6.56 & 5.80 & 7.06  &  Yes  \\
      &  11 & -2.43 & 6416 & 0.66 & 0.56 & 2.06 & -13.21 & -11.79 & 6.46 & 5.70 & 6.96  &  Yes  \\
      &  13 & -2.56 & 6434 & 0.63 & 0.41 & 1.91 & -13.35 & -11.94 & 6.40 & 5.64 & 6.90  &  Yes  \\
\hline
 0.78 &  0  & -2.5  & 5147 & 39.0 & ... & 2.72 & -13.06 & -11.65 & 6.24 & 5.48 & 6.74  &  Yes  \\
      &  8  & -2.83 & 6380 & 0.75 & 0.69 & 2.22 & -13.55 & -12.14 & 6.02 & 5.26 & 6.52  &  Yes  \\
      &  11 & -3.09 & 6514 & 0.61 & 0.43 & 1.88 & -13.89 & -12.48 & 5.89 & 5.13 & 6.39  &  Yes  \\
      &  13 & -3.33 & 6499 & 0.62 & 0.29 & 1.63 & -14.14 & -12.73 & 5.77 & 5.01 & 6.27  &  Yes  \\
\hline
 0.82 &  0  & -0.7  & 4652 & 40.0 & ... & 2.72 & -11.26 & -9.85 & 8.04 & 7.28 & 8.54  &  Yes  \\
      &  8  & -0.8  & 5803 & 0.89 & 1.947 & 2.03 & -11.43 & -10.03 & 7.96 & 7.20 & 8.46  &  Yes  \\
      &  11 & -0.83 & 5877 & 0.77 & 1.82 & 1.93 & -11.50 & -10.07 & 7.94 & 7.18 & 8.43  &  Yes  \\
      &  13 & -0.85 & 5902 & 0.70 & 1.72 & 1.88 & -11.54 & -10.11 & 7.92 & 7.16 & 8.42  &  Yes  \\
\hline
 0.82 &  0  & -1.0  & 4785 & 40.0 & ... & 2.72 & -11.56 & -10.15 & 7.72 & 6.96 & 8.22  &  Yes  \\
      &  8  & -1.13 & 6046 & 0.86 & 1.48 & 2.36 & -11.78 & -10.36 & 7.64 & 6.88 & 8.14  &  Yes  \\
      &  11 & -1.18 & 6109 & 0.77 & 1.26 & 2.27 & -11.87 & -10.45 & 7.60 & 6.84 & 8.10  &  Yes  \\
      &  13 & -1.20 & 6049 & 0.71 & 1.19 & 2.24 & -11.88 & -10.47 & 7.60 & 6.84 & 8.09  &  Yes  \\
\hline
 0.82$^{h}$  &  0  & -1.5  & 4921 & 40.0 & ... & 2.72 & -12.06 & -10.65 & 7.24 & 6.48 & 7.74  &  Yes  \\
       &  8  & -1.76 & 6343 & 0.89 & 0.79 & 2.31 & -12.45 & -11.04 & 7.07 & 6.31 & 7.57  &  Yes  \\
       &  11 & -1.91 & 6385 & 0.80 & 0.54 & 2.11 & -12.64 & -11.23 & 6.99 & 6.23 & 7.49  &  Yes  \\
\hline
 0.82$^{h}$ &  0  & -2.0  & 5061 & 40.0 & ... & 2.72 & -12.56 & -11.15 & 6.74 & 5.98 & 7.24  &  Yes  \\
      &  8  & -2.47 & 6523 & 0.83 & 0.50 & 2.08 & -13.18 & -11.77 & 6.47 & 5.71 & 6.97  &  Yes  \\
      &  11 & -2.75 & 6558 & 0.82 & 0.28 & 1.81 & -13.45 & -12.04 & 6.34 & 5.58 & 6.84  &  Yes  \\
\hline
  0.89$^{h}$ &  0 & -0.70 &  4723 & 40.0 & 1.39 & 2.72 & -11.26 & -9.85 & 8.04 & 7.28 & 8.54 & Yes \\
        &  8 & -0.83 &  6081 & 1.11 & 1.48 & 2.25 & -11.47 & -10.05 & 7.95 & 7.19 & 8.44 & Yes \\
        & 11 & -0.85 &  5984 & 0.91 & 1.40 & 2.19 & -11.50 & -10.08 & 7.93 & 7.17 & 8.43 & Yes \\
\hline
  0.89$^{h}$ &  0 & -1.00 &  4832 & 40.0 & 1.36 & 2.72 & -11.56 & -10.15 & 7.74 & 6.98 & 8.24 & Yes \\
         &  8 & -1.21 &  6299 & 1.09 & 0.87 & 2.33 & -11.88 & -10.46 & 7.60 & 6.84 & 8.10 & Yes \\
         & 11 & -1.01 &  4998 & 0.11 & 2.84 & 0.99 & -12.74 & -10.75 & 7.71 & 7.07 & 8.23 & Yes \\
\enddata
\tablenotetext{}{Notes. The initial Be and B abundances of solar-scaled mixture models
are same as those of the Sun, while those of $\alpha$-enhanced mixture models are determined by
Equations (\ref{beeq}) and (\ref{beq}). }
\tablenotetext{a}{The abundances of C, O, Mg, Si, S, Ar, Ca, and Ti were enhanced according to
\citet{cayr04} and \citet{spit05}. Symbol ``No'' presents models having solar-scaled mixture.}
\tablenotetext{b}{Fully convective model.}
\tablenotetext{c}{ZAMS.}
\tablenotetext{d}{MSTO.}
\tablenotetext{e}{TAMS.}
\tablenotetext{f}{The middle of the SGB.}
\tablenotetext{g}{The base of the RGB.}
\tablenotetext{h}{By 13 Gyr, these models have already evolved past the RGB bump.}
}
\end{deluxetable*}
\end{longrotatetable}

\begin{longrotatetable}
\begin{deluxetable*} {lcccccccccccc}
\bf{
\tablecaption{Parameters of Nonrotating Models.
\label{tab2}}
\tablehead{
\colhead{Mass} & \colhead{Age} & \colhead{[Fe/H]} & \teff{} &\colhead{Velocity} & $T_{\mathrm{BCZ}}$ &\colhead{$A$(Li)} & \colhead{$A$(Be) or} & \colhead{$A$(B) or} & \colhead{$A$(C)} &\colhead{$A$(N)} &\colhead{$A$(O)} & \colhead{$\alpha$-enhanced} \\
\colhead{\dsm{}} & \colhead{Gyr} & & K &\colhead{km s$^{-1}$} & 10$^{6}$ K & & log(Be/H) & log(B/H) &  &  &  &
}
\startdata
  0.65 &  0   & -1.00 &  4741 & 0.0 & ...$^{a}$ & 2.72 & 1.44 & 2.85 & 7.55 & 6.97 & 7.76 & No \\
       & 0.06$^{b}$   & -1.00 &  5051 & 0.0 & 2.12 & 2.65 & 1.44 & 2.85 & 7.55 & 6.97 & 7.76 & No \\
       &  8   & -1.08 &  5189 & 0.0 & 2.06 & 2.46 & 1.25 & 2.66 & 7.47 & 6.89 & 7.68 & No \\
       & 11   & -1.11 &  5259 & 0.0 & 2.01 & 2.38 & 1.18 & 2.59 & 7.44 & 6.86 & 7.65 & No \\
       & 13   & -1.13 &  5311 & 0.0 & 1.98 & 2.34 & 1.14 & 2.55 & 7.42 & 6.84 & 7.63 & No \\
       & 26.6$^{c}$ & -1.25 &  5678 & 0.0 & 1.63 & 2.03 & 0.83 & 2.24 & 7.30 & 6.72 & 7.51 & No \\
       & 27.9$^{d}$ & -1.25 &  5655 & 0.0 & 1.68 & 2.06 & 0.86 & 2.27 & 7.31 & 6.73 & 7.52 & No \\
       & 29.0$^{e}$ & -1.14 &  5443 & 0.0 & 2.02 & 1.67 & 1.02 & 2.59 & 7.41 & 6.83 & 7.62 & No \\
       & 29.6$^{f}$ & -1.07 &  5187 & 0.0 & 2.60 & 1.13 & 0.49 & 2.48 & 7.48 & 6.90 & 7.69 & No \\
\hline
  0.74 &  0 & -1.00 &  4813 & 0.0 & ...  & 2.72 & 1.44 & 2.85 & 7.55 & 6.97 & 7.76 & No \\
       &  8 & -1.12 &  5737 & 0.0 & 1.66 & 2.43 & 1.18 & 2.58 & 7.44 & 6.86 & 7.65 & No \\
       & 11 & -1.16 &  5825 & 0.0 & 1.54 & 2.33 & 1.07 & 2.48 & 7.39 & 6.81 & 7.60 & No \\
       & 13 & -1.19 &  5882 & 0.0 & 1.46 & 2.26 & 1.00 & 2.41 & 7.37 & 6.79 & 7.58 & No \\
\hline
  0.78 &  0 & -1.00 &  4847 & 0.0 & ...  & 2.72 & 1.44 & 2.85 & 7.55 & 6.97 & 7.76  & No \\
       &  8 & -1.15 &  5953 & 0.0 & 1.41 & 2.36 & 1.10 & 2.51 & 7.40 & 6.82 & 7.61  & No \\
       & 11 & -1.20 &  6033 & 0.0 & 1.22 & 2.22 & 0.96 & 2.37 & 7.35 & 6.77 & 7.56  & No \\
       & 13 & -1.25 &  6073 & 0.0 & 1.06 & 2.12 & 0.86 & 2.27 & 7.31 & 6.73 & 7.52  & No \\
\hline
  0.82 &  0         & -1.00 &  4841 & 0.0 & ...  & 2.72 & 1.44 & 2.85 & 7.55 & 6.97 & 7.76  & No \\
       &  0.035$^{b}$ & -1.00 &  5935 & 0.0 & 1.59 & 2.71 & 1.44 & 2.85 & 7.55 & 6.97 & 7.76 & No \\
       &  8 & -1.20 &  6133 & 0.0 & 1.09 & 2.24 & 0.97 & 2.38 & 7.35 & 6.77 & 7.56  & No \\
       & 11$^{c}$   & -1.29 &  6186 & 0.0 & 0.83 & 2.03 & 0.76 & 2.17 & 7.27 & 6.69 & 7.48  & No \\
       & 12.3$^{d}$ & -1.30 &  6104 & 0.0 & 0.80 & 1.99 & 0.72 & 2.13 & 7.25 & 6.67 & 7.46 & No \\
       & 12.97$^{e}$ & -1.13 &  5809 & 0.0 & 1.25 & 2.48 & 1.23 & 2.64 & 7.42 & 6.84 & 7.63 & No \\
       & 13          & -1.13 &  5785 & 0.0 & 1.33 & 2.44 & 1.23 & 2.64 & 7.42 & 6.84 & 7.63  & No \\
       & 13.37$^{f}$ & -1.03 &  5253 & 0.0 & 2.51 & 1.31 & 0.52 & 2.41 & 7.52 & 6.94 & 7.73 & No\\
\hline
  0.74 &  0 & -0.70 &  4661 & 0.0 & 1.56 & 2.72 & -11.26 & -9.85 & 8.04 & 7.28 & 8.54 & Yes \\
       &  8 & -0.78 &  5389 & 0.0 & 2.25 & 2.32 & -11.44 & -10.03 & 7.96 & 7.20 & 8.46 & Yes \\
       & 11 & -0.81 &  5469 & 0.0 & 2.19 & 2.25 & -11.51 & -10.10 & 7.93 & 7.17 & 8.43 & Yes \\
       & 13 & -0.83 &  5522 & 0.0 & 2.13 & 2.20 & -11.55 & -10.14 & 7.91 & 7.15 & 8.41 & Yes \\
\hline
  0.74 &  0 & -1.00 &  4775 & 0.0 & 1.07 & 2.72 & -11.56 & -10.15 & 7.74 & 6.98 & 8.24 & Yes \\
       &  8 & -1.10 &  5653 & 0.0 & 1.93 & 2.44 & -11.79 & -10.38 & 7.64 & 6.88 & 8.14 & Yes \\
       & 11 & -1.14 &  5740 & 0.0 & 1.82 & 2.35 & -11.88 & -10.47 & 7.60 & 6.84 & 8.10 & Yes \\
       & 13 & -1.16 &  5796 & 0.0 & 1.76 & 2.29 & -11.94 & -10.53 & 7.58 & 6.82 & 8.08 & Yes \\
\hline
  0.74 &  0 & -1.50 &  4923 & 0.0 & 1.06 & 2.72 & -12.06 & -10.65 & 7.24 & 6.48 & 7.74 & Yes \\
       &  8 & -1.65 &  5948 & 0.0 & 1.43 & 2.37 & -12.40 & -10.99 & 7.09 & 6.33 & 7.59 & Yes \\
       & 11 & -1.71 &  6038 & 0.0 & 1.26 & 2.22 & -12.55 & -11.14 & 7.03 & 6.27 & 7.53 & Yes \\
       & 13 & -1.75 &  6095 & 0.0 & 1.10 & 2.11 & -12.66 & -11.25 & 6.99 & 6.23 & 7.49 & Yes \\
\hline
  0.74 &  0 & -2.00 &  5140 & 0.0 & 1.49 & 2.72 & -12.56 & -11.15 & 6.74 & 5.98 & 7.24 & Yes \\
       &  8 & -2.20 &  6087 & 0.0 & 1.11 & 2.23 & -13.04 & -11.63 & 6.54 & 5.78 & 7.04 & Yes \\
       & 11 & -2.30 &  6187 & 0.0 & 0.90 & 1.99 & -13.28 & -11.87 & 6.44 & 5.68 & 6.94 & Yes \\
       & 13 & -2.38 &  6257 & 0.0 & 0.74 & 1.79 & -13.49 & -12.08 & 6.36 & 5.60 & 6.86 & Yes \\
\hline
  0.78 &  0 & -0.70 &  4716 & 0.0 & ...  & 2.72 & -11.26 & -9.85 & 8.04 & 7.28 & 8.54 & Yes\\
       &  8 & -0.79 &  5612 & 0.0 & 2.11 & 2.37 & -11.47 & -10.06 & 7.95 & 7.19 & 8.45 & Yes \\
       & 11 & -0.83 &  5692 & 0.0 & 2.03 & 2.29 & -11.54 & -10.13 & 7.91 & 7.15 & 8.41 & Yes \\
       & 13 & -0.85 &  5740 & 0.0 & 1.95 & 2.24 & -11.59 & -10.18 & 7.89 & 7.13 & 8.39 & Yes \\
\hline
  0.78 &  0 & -1.00 &  4822 & 0.0 & 1.64 & 2.72 & -11.56 & -10.15 & 7.72 & 6.96 & 8.22 & Yes \\
       &  8 & -1.13 &  5865 & 0.0 & 1.73 & 2.42 & -11.83 & -10.42 & 7.61 & 6.84 & 8.10 & Yes \\
       & 11 & -1.18 &  5946 & 0.0 & 1.58 & 2.31 & -11.94 & -10.53 & 7.56 & 6.80 & 8.06 & Yes \\
       & 13 & -1.21 &  5987 & 0.0 & 1.44 & 2.24 & -12.01 & -10.60 & 7.53 & 6.77 & 8.03 & Yes \\
\hline
  0.78 &  0 & -1.50 &  4964 & 0.0 & 1.14 & 2.72 & -12.06 & -10.65 & 7.24 & 6.48 & 7.74 & Yes \\
       &  8 & -1.70 &  6155 & 0.0 & 1.09 & 2.23 & -12.54 & -11.13 & 7.04 & 6.28 & 7.54 & Yes \\
       & 11 & -1.80 &  6237 & 0.0 & 0.82 & 1.99 & -12.78 & -11.37 & 6.94 & 6.18 & 7.44 & Yes \\
       & 13 & -1.86 &  6230 & 0.0 & 0.69 & 1.83 & -12.94 & -11.53 & 6.88 & 6.12 & 7.38 & Yes \\
\hline
  0.78 &  0 & -2.00 &  5102 & 0.0 & 1.61 & 2.72 & -12.56 & -11.15 & 6.74 & 5.98 & 7.24 & Yes \\
       &  8 & -2.31 &  6285 & 0.0 & 0.80 & 1.98 & -13.30 & -11.89 & 6.43 & 5.67 & 6.93 & Yes \\
       & 11 & -2.50 &  6394 & 0.0 & 0.52 & 1.46 & -13.82 & -12.41 & 6.24 & 5.48 & 6.74 & Yes \\
       & 13 & -2.65 &  6369 & 0.0 & 0.39 & 1.04 & -14.24 & -12.83 & 6.09 & 5.33 & 6.59 & Yes \\
\hline
  0.78 &  0 & -2.50 &  5123 & 0.0 & 1.14 & 2.72 & -13.06 & -11.65 & 6.24 & 5.48 & 6.74 & Yes \\
       &  8 & -2.91 &  6372 & 0.0 & 0.64 & 1.69 & -14.09 & -12.68 & 5.83 & 5.07 & 6.33 & Yes \\
       & 11 & -3.25 &  6503 & 0.0 & 0.36 & 0.74 & -15.04 & -13.63 & 5.49 & 4.73 & 5.99 & Yes \\
       & 13 & -3.54 &  6387 & 0.0 & 0.28 & -0.06 & -15.84 & -14.43 & 5.20 & 4.44 & 5.70 & Yes \\
\hline
  0.82 &  0 & -0.70 &  4763 & 0.0 & 1.24 & 2.72 & -11.26 & -9.85 & 8.04 & 7.28 & 8.54 & Yes \\
       &  8 & -0.81 &  5808 & 0.0 & 1.92 & 2.38 & -11.50 & -10.09 & 7.93 & 7.17 & 8.43 & Yes\\
       & 11 & -0.84 &  5876 & 0.0 & 1.79 & 2.29 & -11.59 & -10.18 & 7.90 & 7.14 & 8.40 & Yes \\
       & 13 & -0.87 &  5890 & 0.0 & 1.71 & 2.24 & -11.64 & -10.23 & 7.87 & 7.11 & 8.37 & Yes \\
\hline
  0.82 &  0 & -1.00 &  4866 & 0.0 & 1.22 & 2.72 & -11.56 & -10.15 & 7.74 & 6.98 & 8.24 & Yes\\
       &  8 & -1.15 &  6049 & 0.0 & 1.46 & 2.36 & -11.90 & -10.49 & 7.59 & 6.83 & 8.09 & Yes \\
       & 11 & -1.20 &  6098 & 0.0 & 1.21 & 2.22 & -12.03 & -10.62 & 7.54 & 6.78 & 8.04 & Yes \\
       & 13 & -1.21 &  5976 & 0.0 & 1.24 & 2.21 & -12.04 & -10.63 & 7.53 & 6.77 & 8.03 & Yes \\
\hline
  0.82 &  0 & -1.50 &  5004 & 0.0 & 1.21 & 2.72 & -12.06 & -10.65 & 7.24 & 6.48 & 7.74 & Yes \\
       &  8 & -1.82 &  6333 & 0.0 & 0.73 & 1.95 & -12.83 & -11.42 & 6.92 & 6.16 & 7.42 & Yes  \\
       & 11 & -1.98 &  6326 & 0.0 & 0.50 & 1.51 & -13.26 & -11.85 & 6.76 & 6.00 & 7.26 & Yes \\
\hline
  0.82 &  0 & -2.00 &  5087 & 0.0 & 1.28 & 2.72 & -12.56 & -11.15 & 6.74 & 5.98 & 7.24 & Yes \\
       &  8 & -2.60 &  6492 & 0.0 & 0.45 & 1.16 & -14.12 & -12.71 & 6.14 & 5.38 & 6.64 & Yes\\
       & 11 & -3.17 &  6482 & 0.0 & 0.25 & -0.49 & -15.77 & -14.36 & 5.57 & 4.81 & 6.07 & Yes \\
\hline
  0.89 &  0 & -0.70 &  4756 & 0.0 & 1.41 & 2.72 & -11.26 & -9.85 & 8.04 & 7.28 & 8.54 & Yes \\
       &  8 & -0.84 &  6067 & 0.0 & 1.46 & 2.33 & -11.59 & -10.18 & 7.90 & 7.14 & 8.40 & Yes \\
       & 11 & -0.86 &  5944 & 0.0 & 1.41 & 2.30 & -11.62 & -10.21 & 7.88 & 7.12 & 8.38 & Yes \\
\hline
  0.89 &  0 & -1.00 &  4952 & 0.0 & 1.36 & 2.72 & -11.56 & -10.15 & 7.74 & 6.98 & 8.24 & Yes \\
       &  8 & -1.26 &  6276 & 0.0 & 0.82 & 2.10 & -12.17 & -10.76 & 7.48 & 6.72 & 7.98 & Yes \\
\enddata
\tablenotetext{}{Notes. the initial Be and B abundances of models without $\alpha$-enhancement
are same as those of the Sun, while those of models with $\alpha$-enhancement are determined by
Equations (\ref{beeq}) and (\ref{beq}). }
\tablenotetext{a}{Fully convective model.}
\tablenotetext{b}{ZAMS.}
\tablenotetext{c}{MSTO.}
\tablenotetext{d}{TAMS.}
\tablenotetext{e}{The middle of the SGB.}
\tablenotetext{f}{The base of the RGB.}
}
\end{deluxetable*}
\end{longrotatetable}

\end{document}